\documentclass[prd,preprint,eqsecnum,nofootinbib,amsmath,amssymb,
               showpacs,tightenlines,dvips]{revtex4}
\usepackage{graphicx}
\usepackage{bm}
\input epsf

\begin {document}


\def\half{\tfrac12}
\def\eps{\epsilon}
\def\p{{\bm p}}
\def\q{{\bm q}}
\def\k{{\bm k}}
\def\l{{\bm l}}
\def\v{{\bm v}}
\def\s{{\bm s}}
\def\S{{\cal S}}
\def\T{{\cal T}}
\def\E{{\cal E}}


\vspace*{-1cm}
\begin{flushright}
{INT-PUB 06--12}
\end{flushright}
\vspace*{1cm}

\title
    {
     Next-to-next-to-leading-order \boldmath$\epsilon$ expansion for 
     a Fermi gas
     at infinite scattering length
    }

\author{Peter Arnold}
\affiliation
    {%
    Department of Physics,
    University of Virginia, Box 400714,
    Charlottesville, Virginia 22904-4714, USA
    }%
\author{Joaqu\'in E. Drut and Dam Thanh Son}
\affiliation
    {%
    Institute for Nuclear Theory,
    University of Washington,
    Seattle, Washington 98195-1550, USA
    }%

\date {August 2006}

\begin {abstract}%
    {%
      We extend previous work on applying the $\epsilon$ expansion to
      universal properties of a cold, dilute Fermi gas in the unitary
      regime of infinite scattering length.
      We compute the ratio $\xi=\mu/\epsilon_{\rm F}$
      of chemical potential to ideal gas Fermi energy to
      next-to-next-to-leading order (NNLO) in $\epsilon=4-d$, where $d$ is
      the number of spatial dimensions.  We also explore the nature of
      corrections from the order after NNLO.
    }%
\end {abstract}
 
\maketitle
\thispagestyle {empty}


\section {Introduction and Results}
\label{sec:intro}

For a number of years, it has been a challenge to compute the properties
of a dilute Fermi gas with infinite scattering length \cite{unitary}.  
This is known
as the unitary regime and is relevant to cold systems for which the
interparticle separation is very small compared to the two-particle
scattering length $a$, but very large compared to any other distance scales
$r$ characterizing two-body interactions:
$a \gg n^{-1/3} \gg r$, where $n$ is the density.
In the unitary regime, the only dimensionful parameter is the number
density $n$, and so all physical quantities in this limit should be
determined by dimensional analysis and universal constants of
proportionality.
Dimensionless ratios will be universal \cite{universal}.
There has been much interest in recent years in attempting to compute
such universal constants of the unitary regime (see, e.g., 
Refs.\ \cite{BakerHeiselberg,GFMC} and references therein).

The problem is nonperturbative in three spatial dimensions.
However, inspired by earlier work of Nussinov and Nussinov
\cite{Nussinov}
on the behavior of
the unitary regime as a function of
spatial dimension $d$,
it was recently realized \cite{Nishida&Son}
that, with an appropriately formulated
perturbation theory,
a perturbative solution is possible in $d{=}4{-}\eps$ spatial dimensions
when $\eps \ll 1$.
Results can be expressed as an asymptotic series in $\eps$, analogous to
the $\eps$ expansion methods
that have been used with great success for 30 years
to determine critical exponents in a variety of second-order phase
transitions.  One may then extrapolate to the case of three dimensions,
$\eps=1$.  For the case of cold, dilute Fermi gases at infinite scattering
length, it is found \cite{Nishida&Son} that
\begin {align}
   \xi \equiv \frac{\mu}{\eps_{\rm F}}
   &= \half \eps^{\eps/2d} \left[ \eps^{3/2}
        - 0.0492 \, \eps^{5/2}
        + O(\eps^{7/2})
      \right]
   \notag\\
   &= \half \eps^{3/2}
       + \tfrac1{16}\eps^{5/2}\ln\eps
       - 0.0246 \, \eps^{5/2}
       + \cdots ,
\label{eq:xiNS}
\\
   \frac{\eps_0}{\mu}
   &= 2 + O(\eps) ,
\\
   \frac{\Delta}{\mu}
   &= \frac{2}{\eps} - 0.691 + O(\eps) ,
\end {align}
where $\mu$ is the chemical potential, $\Delta$ is the gap for
fermionic excitations, and $\eps_0$ is the value of $p^2/2m$
for fermionic excitations with the minimum energy $\Delta$.
The Fermi energy $\eps_{\rm F}$ is defined as the Fermi energy
of an ideal Fermi gas with the same density $n$ as the
strongly interacting gas under consideration.

The ratio $\xi$ can be equivalently expressed as an energy density 
ratio\footnote{This can be proven simply by using thermodynamics 
and scaling at unitarity \cite{universal}.}
\begin {equation}
   \xi = \frac{{\cal E}}{{\cal E}_0} \,,
\end {equation}
where ${\cal E}$ and ${\cal E}_0$ are the
energy densities in the interacting and noninteracting cases, respectively,
at equal number density $n$.
For experimental relevance, $\xi$ can also
be expressed as \cite{release energy,exp_Duke2}

\begin {equation}
   \xi = \left(\frac{E^{\rm rel}}{E^{\rm rel}_0}\right)^2
         \qquad \mbox{(harmonic trap)}
\end {equation}
for a system in a harmonic trap (in the limit of an
arbitrarily wide trap),
where the ``release'' energies
$E^{\rm rel}$ and $E^{\rm rel}_0$ are the total system energies in
the interacting and noninteracting cases at equal total particle
number $N$.




Historically, in the simplest application of the $\eps$ expansion,
one computes the
first two or three terms in the $\eps$ expansion and sees if they are
reasonably well behaved for $\eps{=}1$, which corresponds to three spatial
dimensions.  Using this method, the above expansions would give the
estimates \cite{Nishida&Son}
\begin {equation}
   \xi \simeq 0.475,
   \qquad
   \frac{\eps_0}{\mu} \simeq 2,
   \qquad
   \frac{\Delta}{\mu} \simeq 1.31
   \,.
\label {eq:NLO}
\end {equation}
In this method, it is important to quit when one is ahead: Because the
$\eps$ expansion is asymptotic, higher-order terms eventually grow.  The
simple procedure is to stop including higher-order terms when this
happens.  In more sophisticated applications of the $\eps$ expansion,
however, critical exponents have been determined fairly precisely for
some phase transitions by combining high-order $\eps$ expansions,
information about the large-order asymptotic behavior, and knowledge of
the behavior at or near lower dimensions such as $d{=}2$, to fit results
as a function of dimension using Borel-Pad\'e approximations
\cite{BGZ2,sophisticated eps,Berviller,current exponents}.%
\footnote{
  For a textbook overview, see chapters 28 and 41 of
  Ref.\ \cite{ZinnJustinBook}.
}
A similar procedure has recently been carried out for the ratio
$\xi = \mu/\eps_{\rm F}$
in Ref.\ \cite{Nishida&Son2} using next-to-leading-order results
for the $\epsilon$ expansion about both four and two spatial dimensions
(except that the high-order asymptotic behavior of the $\epsilon$
expansion about four dimensions is currently unknown).

The goal of the current paper is to take the step of computing the
next term in the $\epsilon$ expansion about four spatial dimensions
for $\xi$ at zero temperature, and to learn something about
the analytic structure in $\epsilon$ by showing that new, non-trivial
logarithms of $\epsilon$ appear at yet higher orders.
We find%
\begin {equation}
   \xi
   = \half \eps^{\eps/2d} \left[ \eps^{3/2}
        - 0.04916 \, \eps^{5/2}
        - 0.95961 \, \eps^{7/2}
        - \tfrac38 \, \eps^{9/2} \ln\eps
        + O(\eps^{9/2})
     \right].
\label {eq:xi}
\end {equation}
One may expand
\begin {equation}
  \eps^{\eps/2d} =
  1 + \tfrac18 \, \eps \ln\eps
    + \eps^2 \left( \tfrac1{128} \ln^2\eps + \tfrac1{32} \ln\eps \right)
    + O(\eps^3 \ln^3\eps)
\end{equation}
if desired, as in Eq.\ (\ref{eq:xiNS}).
If one intends to naively set $\eps{=}1$, this expansion is unnecessary
since $\ln\eps$ then vanishes and $\eps^{\eps/2d} = 1$ order
by order in $\eps$.
We will comment on the large relative size of the 
next-to-next-to-leading-order (NNLO) correction
at the end of the paper and discuss there the implications of our result for
Borel-Pad\'e extrapolations to $d{=}3$.

In the remainder of this paper, we explain our calculation of $\xi$
to NNLO in the $\eps$ expansion.
In the next section, we briefly review the formalism and diagrammatic
rules developed in Ref.\ \cite{Nishida&Son} for applying the
$\eps$ expansion to this problem.
At the beginning of Sec.\ \ref{sec:NNLO}, we display the diagrams that need
to be evaluated to push the calculation of the effective potential
(which is later used to determine $\xi$) to
NNLO in $\eps$.
Because the efficient evaluation of some diagrams is
challenging, we will then take the time to explain our methods in detail.
Results for all the diagrams are summarized in Appendix \ref{app:summary}.
In Sec.\ \ref{sec:xi}, we put everything together to determine
$\xi$ as in Eq.\ (\ref{eq:xi}).
Finally, in Sec.\ \ref{sec:beyond NNLO}, we explain how the
diagrammatic
$\eps$ power counting of Ref.\ \cite{Nishida&Son}
would break down, due to infrared
issues, if we proceeded to yet one higher order in $\eps$ than the
calculation reported in this paper.
We then show how the proper power counting
of $\eps$ can be restored.
Finally, we discuss the implications of our result to extrapolating
the value of $\xi$ to $d{=}3$
in Sec.\ \ref{sec:extrapolate}.


\section{Review}
\label {sec:review}

We will generally follow the conventions and diagrammatic methods of
Ref.\ \cite{Nishida&Son}, which we review here.
One starts with the Lagrangian
\begin{equation}\label{L}
   {\cal L} =
   \Psi^\dagger \left( i \partial_t + \frac{\nabla^2}{2m} \, \sigma_3 \right)
                \Psi
   + \mu \Psi^\dagger \sigma_3 \Psi
   - \frac{1}{c_0} \phi^* \phi
   + \Psi^\dagger \sigma_+ \Psi \phi
   + \Psi^\dagger \sigma_- \Psi \phi^* ,
\end{equation}
where $\phi$ is a Hubbard-Stratonovich field,
$\Psi = (\psi_\uparrow,\psi_\downarrow^\dagger)^\top$ is
a two-component Nambu-Gor'kov field,
$\sigma_\pm = \half(\sigma_1 \pm i \sigma_2)$,
and $\sigma_{1,2,3}$ are the Pauli matrices.
The constant $c_0$ determines the scattering length.  In three spatial
dimensions,
\begin{equation}\label{ac0}
  \frac m{4\pi a} = - \frac1{c_0} + \int\!\frac{d^3p}{(2\pi)^3}\,
  \frac m{p^2}\,.
\end{equation}
The integral on the right hand side of Eq.~(\ref{ac0}) is ultraviolet 
divergent (and is so in any dimension $d\ge2$).  In a physical 
system there is always an upper momentum cutoff, for example,
the inverse of the range of potential.  However, if the system
is insensitive
to the cutoff, then $c_0$ should always appear in physical observables
in the same combination as
in Eq.~(\ref{ac0}), so that observables can be expressed in terms
of the scattering length.

Technically, it is cumbersome to carry the integral in Eq.~(\ref{ac0})
across our formulas, so we will use dimensional regularization.  The
technical advantage of this regularization scheme is that the integral
$\int\!d^dp/p^2$ vanishes in any dimension, and so the connection
between $c_0$ and $a$ becomes very simple.  In particular, infinite
scattering length corresponds to $c_0=\infty$, so that there is no term
quadratic in $\phi$ in Eq.~(\ref{L}).  As long
as the physics is insensitive to short distances, dimensional
regularization will give the same result as other regularization
schemes such as a momentum cutoff.%
%
\footnote{%
See Ref. \cite{PeskinBook} for
a standard textbook treatment of dimensional regularization, with emphasis
in high-energy physics.
For condensed matter applications 
see, for instance, the textbook
treatment in Ref. \cite{ZinnJustinBook}.
For a few examples of use in the theory of cold, dilute atomic gases,
see Ref.\ \cite{dimreg}.
}

\begin{figure}[t]
\includegraphics[scale=0.50]{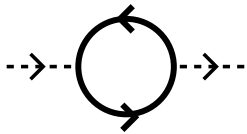}
\caption{%
    \label{fig:scalar}
    The one-loop scalar self-energy $\Pi$, where solid lines represent
    Nambu-Gor'kov fermion fields $\Psi$ and dashed lines represent the
    scalar field variation $\varphi$.
    (There is a similar diagram where one of the external
    scalar arrows is reversed,
    mixing $\varphi$ with $\varphi^*$, which we will discuss later and
    denote $\widetilde\Pi$.)
    }
\end{figure}

Following Ref.\ \cite{Nishida&Son}, we expand $\phi$ about its
superfluid expectation value $\langle \phi \rangle \equiv \phi_0$ as
\begin {equation}
   \phi = \phi_0 + g\varphi ,
\end {equation}
where $g$ is chosen to give the dynamics of $\varphi$ a conventional
normalization at leading order in $\eps$.  In particular, if one computes
the small-momentum expansion of the $\varphi$ self-energy $\Pi$ of
Fig.\ \ref{fig:scalar} to leading order in $\eps$, one finds
\begin {equation}\label{Pi0}
   \Pi(p_0,\p) = \Pi(0) 
               + \frac{g^2 m^2}{8\pi^2\eps} \left(-p_0 + \frac{p^2}{4m}\right)
                   [1 + O(\eps)]
               + O(p_0^2,p^4) .
\end {equation}
The choice $g^2m^2 = 8 \pi^2 \eps [1 + O(\eps)]$ will then make the
momentum dependence above into a conventionally normalized kinetic
term for a nonrelativistic particle of mass $M_\varphi = 2m$
for $d=4$.
Reference \cite{Nishida&Son} found it convenient to define
\begin {equation}
   g^2 \equiv \frac{8\pi^2\eps}{m^2}
            \left(\frac{m\phi_0}{2\pi}\right)^{\eps/2} .
\label {eq:g2}
\end {equation}

\begin{figure}[t]
\includegraphics[scale=0.50]{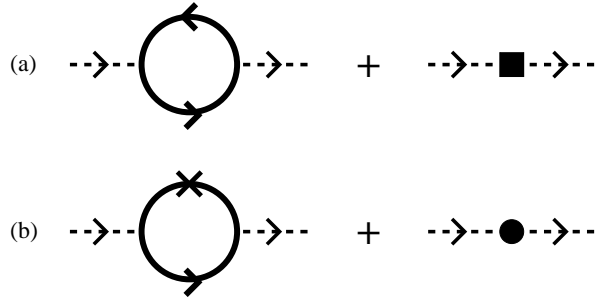}
\caption{%
    The rule for using combining divergent subdiagrams with
    counterterms from ${\cal L}_2$
    to achieve a simple perturbative expansion in
    $\eps$ \cite{Nishida&Son}.
    \label{fig:counterterms}
    }
\end{figure}

\begin{figure}[t]
\includegraphics[scale=0.50]{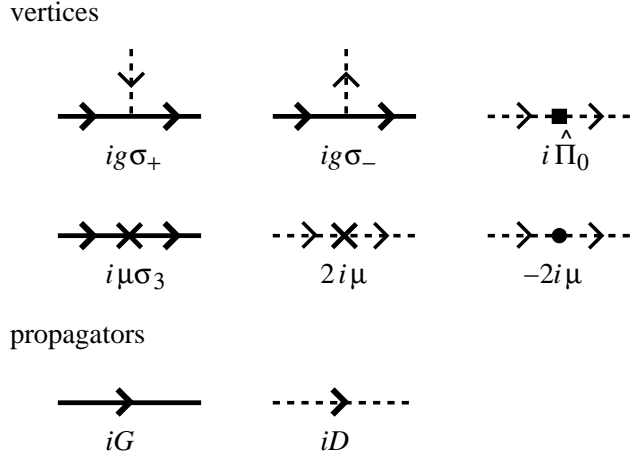}
\caption{%
    \label{fig:Feynman}
    Feynman rules from Eqs.\ (\ref{eq:L}) \cite{Nishida&Son}.
    }
\end{figure}

Reference \cite{Nishida&Son} then reorganized the Lagrangian in the
case of infinite scattering length as a sum
${\cal L} = {\cal L}_0 + {\cal L}_1 + {\cal L}_2$ corresponding
to an unperturbed
Lagrangian ${\cal L}_0$ of a free fermion field $\Psi$ and a free
scalar field $\varphi$, plus perturbations ${\cal L}_1 + {\cal L}_2$:
\begin{subequations}\label {eq:L}
\begin {align}
  {\cal L}_0 &=
    \Psi^\dagger \left( i \partial_t + \frac{\sigma_3 \nabla^2}{2m}
         + \sigma_+ \phi_0 + \sigma_- \phi_0 \right) \Psi
    + \varphi^* \left( i \partial_t + \frac{\nabla^2}{4m} \right) \varphi,
\\
  {\cal L}_1 &=
    g \Psi^\dagger \sigma_+ \Psi \varphi
    + g \Psi^\dagger \sigma_- \Psi \varphi^*
    + \mu \Psi^\dagger \sigma_3 \Psi
    + 2\mu\varphi^*\varphi ,
\\
  {\cal L}_2 &=
    - \varphi^* \left( i \partial_t + \frac{\nabla^2}{4m} \right) \varphi
    - 2 \mu \varphi^* \varphi .
\end {align}
\end{subequations}
Here, ${\cal L}_1$ can be thought of as the interaction terms.
As explained in Ref.\ \cite{Nishida&Son}, ${\cal L}_2$ should be
employed as counterterms to the one-loop diagrams shown in
Fig.\ \ref{fig:counterterms}.%
\footnote{
  The split-up (\ref{eq:L}) might appear unconventional, but the basic idea
  behind it, as in many other occasions in condensed matter physics, is
  to resum divergent graphs.  The graphs that needs to be resummed in our 
  case are multiple insertions of the fermion loop (Fig.~\ref{fig:scalar})
  into the $\varphi$ propagator.  One could, in principle, formulate a set 
  of Feynman rules where the $\varphi$ propagator is the inverse of 
  the fermion loop,
  and fermion loop insertion into the scalar propagator is forbidden by
  hand.  However, the resulting $\varphi$ propagator would be a very 
  complicated function of momentum and chemical potential.  For practical 
  calculations it is 
  much more efficient to give the $\varphi$ propagator a simpler
  form, equal to the inverse of the leading $1/\epsilon$ piece of the 
  fermion loop [Eq.~(\ref{Pi0})], and have it corrected in higher loops.
  To do that at the formal level, the split-up (\ref{eq:L})
  is introduced.
}
%
Feynman rules are shown in
Fig.\ \ref{fig:Feynman}, where
\begin {equation}
   \hat\Pi_0 \equiv -p_0 + \frac{p^2}{4m} \,.
\label {eq:hatPi}
\end {equation}
(This is our one deviation from the notation of Ref.\
\cite{Nishida&Son}.  We find it convenient to define our
$\hat\Pi_0$ as the negative of their $\Pi_0$.
We have added the hat over $\Pi$ to distinguish it and avoid
notational confusion on this point.)
The unperturbed propagators generated by ${\cal L}_0$ are
\begin {equation}
   G(p_0,\p) =
   \begin{pmatrix}
      p_0 - \eps_\p + i\varepsilon & \phi_0 \\
      \phi_0 & p_0 + \eps_\p - i \varepsilon
   \end{pmatrix}^{-1}
   = \frac{1}{p_0^2-E_\p^2+i\varepsilon}
   \begin {pmatrix}
      p_0 + \eps_\p & -\phi_0 \\
      -\phi_0 & p_0 - \eps_\p
   \end {pmatrix}
\label {eq:G}
\end {equation}
and
\begin {equation}
  D(p_0,\p) = \frac{1}{p_0 - \half \eps_\p + i \varepsilon} ,
\label {eq:D}
\end {equation}
where
\begin {equation}
   \eps_\p \equiv \frac{p^2}{2m} ,
   \qquad
   E_\p \equiv (\eps_\p^2+\phi_0^2)^{1/2} .
\end {equation}

By analyzing the effective potential $V(\phi_0)$ for the expectation
$\phi_0$, the minimum is found at $\phi_0 \sim \mu/\eps$,
or equivalently $\mu \sim \eps\phi_0$.  The
correspondence between the diagrammatic expansion and the $\eps$
expansion can then be codified
by treating $\phi_0$ as $O(\eps^0)$ and each insertion
of $\mu$ or $g^2$ [see Eq.\ (\ref{eq:g2})] as $O(\eps)$.  With one
exception,
this identification
gives the relative importance of each diagram of
the effective potential to
NNLO
in the $\eps$ expansion (the order relevant for
the current calculation), provided one uses counterterms according to
Fig.\ \ref{fig:counterterms}.  The one exception is the one-loop
fermion diagram with a single $\mu$ insertion, shown in
Fig.\ \ref{fig:diags1}(b), which produces a $1/\eps$ that
compensates for the $\mu$ and which is not canceled by
any counterterm diagram.

A word of review is in order regarding the nature of the problem above
and below four dimensions.
In $d=4-\epsilon$, the corrections to mean-field theory for the unitary 
Fermi gas are controlled by powers of $\epsilon$.
For $d>4$, on the 
other hand, the problem becomes qualitatively different, as the 
short distance scale $R$ (the range of the potential) renders the 
problem non-universal, and ill defined as $R\rightarrow0$. This is easy to see 
if one solves the two-body problem in arbitrary dimension, as done 
by Nussinov and Nussinov \cite{Nussinov}.
[Precisely at $d=4$, mean-field theory 
acquires logarithmically small corrections of order $1/\log(L/R)$, where
$L \sim n^{-1/3}$ is the typical particle separation.%
\footnote{
  One can understand this logarithm in the spirit of Nussinov and
  Nussinov by recalling that the unitary limit corresponds to the
  presence of a zero-energy bound state, whose wave function will be
  $\Psi \propto r^{2-d}$ outside of the range $R$ of the potential.
  The normalization integral
  $\int d^dr |\Psi|^2 \sim \int d^d r \> r^{4-2d}$
  for the total
  probability is UV convergent in $d=4-\eps$ dimensions but UV
  logarithmically divergent in $d=4$ dimensions, where it introduces a
  logarithmic dependence on $R$.  More technically, if one follows
  the $d=4-\eps$ derivations of Nishida and Son, briefly reviewed
  here, the source of the small parameters $g^2 \sim \eps$ and
  $\mu/\phi_0 \sim \eps$ of the expansion about mean-field theory
  come from logarithmically divergent (in $d=4$) integrals:
  respectively, the $1/\eps$ in the self-energy (\ref{Pi0}) above 
  [from Fig. 1] and
  in the potential (\ref{Vcombined}) [from Fig. 4(b)].
  These integrals $\int d^dp / p^4$ are momentum-space versions
  of $\int d^d r \> r^{4-2d}$.
  Imagine roughly cutting off the integrals
  in the ultraviolet at the scale
  $r \sim R$ and $p \sim 1/R$, where the effective theory breaks down
  and one would need a treatment of the details of the two-body
  potential.
  In the calculations of Nishida and Son and in this paper, the infrared
  is cut off by the distance scale $s \sim (m\phi_0)^{-1/2}$ associated
  with the condensate $\phi_0$.  In $d=4$, we then see that
  the role of $1/\eps$ is replaced by
  $\ln(s/R)$.
  (The resulting solution for $\phi_0$ will relate $s$ and $L$ by a
  power of this logarithm, and so $\ln(s/R) \sim \ln(L/R)$
  up to corrections suppressed by inverse powers of the logarithm.)
  We also learn that, in
  $d=4-\eps$ dimensions with $\eps$ small, $R$ must be exponentially tiny
  in order to be in the universal regime.  Specifically, in order
  for $R$ not to significantly affect $\int d^d p / p^4$, one must
  have $R \ll e^{-1/\eps} s \sim \eps^{1/4} e^{-1/\eps} L$.
}%
]


\section{Evaluating the Next-to-next-to-leading-order Potential}
\label {sec:NNLO}

\subsection {The diagrams}

\begin{figure}[t]
\includegraphics[scale=0.50]{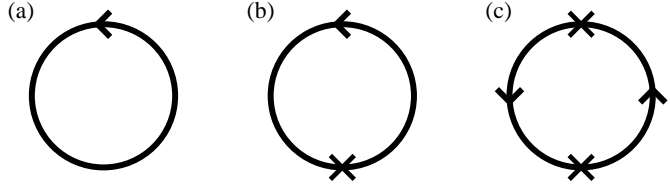}
\caption{%
    \label{fig:diags1}
    One-loop diagrams through $O(\eps^2)$.
    Scalar loops are not shown since these vanish due to the
    retarded nature of the propagators
    We will use the notation
    (a) $V_1^{(0)}$,
    (b) $V_1^{(\mu)}$, and
    (c) $V_1^{(\mu\mu)}$.
    }
\end{figure}

\begin{figure}[t]
\includegraphics[scale=0.50]{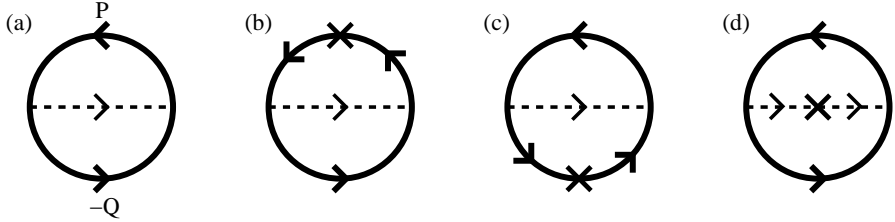}
\caption{%
    \label{fig:diags2}
    Two-loop diagrams.
    The counter-term diagrams for (a--c)---a single scalar loop with an
    appropriate ${\cal L}_2$ counterterm---are not shown because they
    vanish due to the retarded nature of the scalar propagator.
    Our notation is
    (a) $V_2^{(0)}$, and
    (b--d) $V_2^{(\mu)}$.
    }
\end{figure}

\begin{figure}[t]
\includegraphics[scale=0.50]{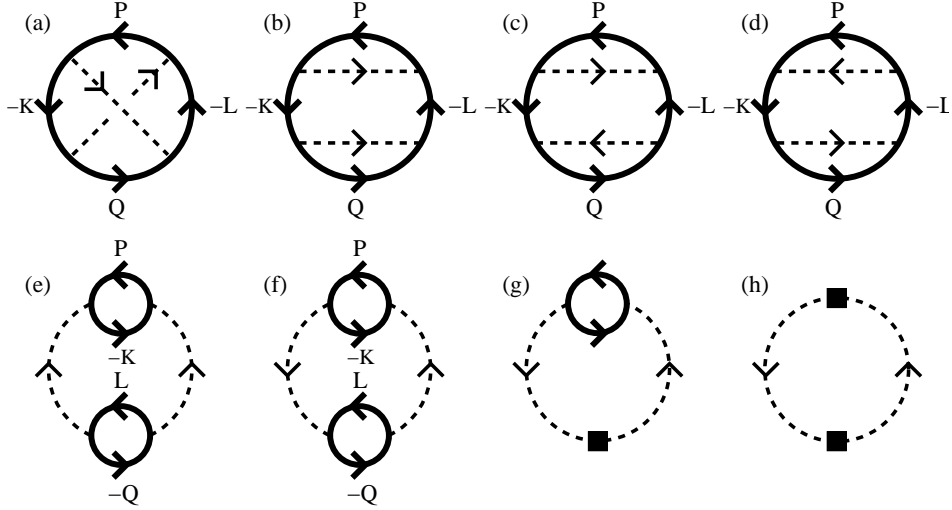}
\caption{%
    \label{fig:diags3}
    Three-loop diagrams (including counterterms to bosonic self-energies). 
    (h) vanishes due to the retarded nature of the scalar
    propagator, but we find it useful to include to make more obvious the
    cancellation of UV divergences.
    Our notation is
    (a) $V_3^{(\otimes)}$,
    (b--d) $V_3^{(\Sigma\Sigma)}$,
    (e) $V_3^{(\widetilde\Pi\widetilde\Pi)}$, and
    (f--h) $V_2^{(\Pi\Pi)}$.
    }
\end{figure}

\begin{figure}[t]
\includegraphics[scale=0.50]{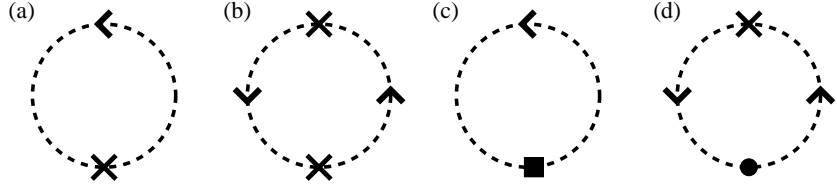}
\caption{%
    \label{fig:vanish}
    Examples of diagrams which vanish because of the purely retarded
    nature of the scalar propagator $D(P)$.
    }
\end{figure}

Figures \ref{fig:diags1}--\ref{fig:diags3} show the nontrivial diagrams
that determine the effective potential $V(\phi_0)$ through NNLO in $\eps$,
together with our conventions for labeling momenta.
In the figure captions, we explain the notation we will use for
the contribution of various classes of these diagrams to the effective
potential.
We have generally not
included various one-loop scalar diagrams which vanish simply because of the
retarded nature of the scalar propagator $D$ of Eq.\ (\ref{eq:D}), such
as those shown in Fig.\ \ref{fig:vanish}.%
\footnote{
  Naively, these diagrams can be seen to vanish by closing the loop frequency
  integration in the upper half plane, which contains no poles.  There
  is a technical caveat, however, in that the contribution from the
  semicircle at infinity cannot be ignored in all cases.  This can
  give $\phi_0$-independent contributions to the effective potential
  which vanish in dimensional regularization and which in other schemes
  correspond to
  operator ordering issues, such as whether $\mu$
  multiplies the Wigner-ordered number operator
  $\half (a^\dagger a + a a^\dagger)$ (corresponding to a naive
  application of the Feynman rules) or the correct normal-ordered
  operator $a^\dagger a$.
  \label {foot:Cinf}
}

\subsection {Cross diagram}

As our first example, we will begin by discussing how to efficiently
evaluate the contribution $V_3^{(\otimes)}(\phi_0)$ of the cross diagram
of Fig.\ \ref{fig:diags3}(a) to the full effective potential $V(\phi_0)$
at $O(\eps^2)$.  This is a three-loop diagram and its leading contribution
is already $O(\eps^2)$ because of the four explicit factors of the
coupling $g$.  We may therefore ignore any other $\eps$ dependence at
the order of interest and so can evaluate the loop integrals for exactly
$d{=}4$.  (For this diagram, the loop integrals converge for
$d{=}4$, which we will see explicitly.)  Our basic approach will
be to evaluate all of the frequency integrals.  One can then scale
all dimensionful parameters out of the remaining momentum integrals.
We then perform the dimensionless momentum integrals numerically.

Figure \ref{fig:diags3}(a) gives
\begin {equation}
  -i V_3^{(\otimes)}
  =
  - (ig)^4 \int_{PQK} i G_{11}(P) \, i G_{22}(Q) \, i G_{21}(-K)
  \, i G_{12}(P{+}Q{+}K) \, i D(P{+}K) \, i D(-Q{-}K) ,
\end {equation}
where the overall minus sign on the right-hand side is for the fermion loop.
We will use capital letters $P$ to stand for $(p_0,\p)$, where
$p_0$ is frequency and $\p$ is spatial momentum,
and we will use the short-hand notations
\begin {equation}
   \int_P \cdots \equiv
      \int \frac{d p_0}{2\pi} \> \frac{d^d p}{(2\pi)^d} \cdots ,
   \qquad
   \int_\p \cdots \equiv
      \int \frac{d^d p}{(2\pi)^d} \cdots .
\end {equation}
Since $G_{21}=G_{12}$, this is the same
as
\begin {equation}
  V_3^{(\otimes)}
  =
  \frac{g^4}{i^3} \int_{PQK} G_{11}(P) \, G_{22}(Q) \, G_{12}(-K)
  \, G_{12}(P{+}Q{+}K) \, D(P{+}K) \, D(-Q{-}K) .
\label {eq:V3cross0}
\end {equation}


\subsubsection {The frequency integrals}

One could simply use the expressions (\ref{eq:G}) and (\ref{eq:D}) for
the propagators and now do the three frequency integrals
($p_0$, $q_0$, $k_0$) by brute force.  This approach is tedious
and yields complicated expressions with many terms requiring significant effort
to simplify.  It also naturally produces terms with energy denominators such as
$(E_\q - E_\k + \half \eps_{\q+\k})^{-1}$ which look like they produce
singularities for certain momenta (e.g., $\q{=}-\k$ in this
example), but all such singularities turn out to cancel between
different terms in the final result.

\begin{figure}[t]
\includegraphics[scale=0.50]{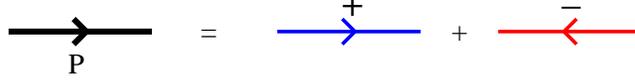}
\caption{%
    \label{fig:Gsplit}
    Split of fermion propagators into retarded and advanced terms
    according to Eq.~(\ref{eq:Gexpand}).
    }
\end{figure}

There is, however, a method for carrying out the frequency integration
which directly produces much tidier results.
The first step is to rewrite
\begin{subequations}
\label{eq:Gexpand}
\begin {eqnarray}
   G_{11}(P) &=& \frac{p_0+\eps_\p}{p_0^2 - E_\p^2 + i\varepsilon}
   = \frac{1}{2E_\p} \left[
        \frac{E_\p+\eps_\p}{(p_0 - E_\p + i\varepsilon)}
      + \frac{E_\p-\eps_\p}{(p_0 + E_\p - i\varepsilon)}
     \right] ,
\\
   G_{22}(Q) &=& \frac{q_0-\eps_\q}{q_0^2 - E_\q^2 + i\varepsilon}
   = \frac{1}{2E_\q} \left[
        \frac{E_\q-\eps_\q}{(q_0 - E_\q + i\varepsilon)}
      + \frac{E_\q+\eps_\q}{(q_0 + E_\q - i\varepsilon)}
     \right] ,
\\
   G_{12}(S) &=& \frac{-\phi_0}{s_0^2 - E_\s^2 + i\varepsilon}
   = \frac{-\phi_0}{2E_\s} \left[
          \frac{1}{(s_0 - E_\s + i\varepsilon)}
        - \frac{1}{(s_0 + E_\s - i\varepsilon)}
     \right] ,
\end {eqnarray}
\end {subequations}
i.e., to decompose the propagators into retarded and advanced parts.
The integrand in Eq.\ (\ref{eq:V3cross0}) then splits into $2^4 = 16$
different terms.  We shall see that many of these terms
trivially vanish, and others are related by symmetry.
It is useful to give a graphical depiction of these different terms
by schematically rewriting Eqs.\ (\ref{eq:Gexpand}) as
in Fig.\ \ref{fig:Gsplit}.  The first and second terms on the right of this
figure denotes the
$(p_0-E_\p+i\varepsilon)^{-1}$ 
and
$(p_0-E_\p-i\varepsilon)^{-1}$
terms on the right-hand side of Eqs.\ (\ref{eq:Gexpand}).
The $+$ and $-$ signs in Fig.\ \ref{fig:Gsplit} denote the sign of
$i\varepsilon$, and the direction of the arrows on the right-hand
side of Fig.\ \ref{fig:Gsplit} correspondingly represent the flow of time
(forward for a retarded propagator, backward for an advanced one).

\begin{figure}[t]
\includegraphics[scale=0.30]{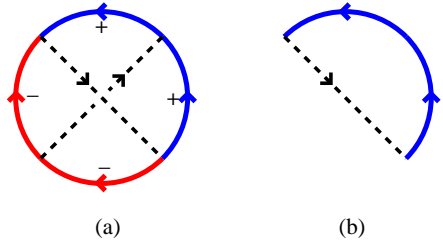}
\caption{%
    \label{fig:term}
    (a) One of the 16 terms generated from Fig.\ \ref{fig:diags3}a
    by the expansions (\ref{eq:Gexpand}).
    (b) A vanishing loop in this diagram.
    }
\end{figure}

\begin{figure}[t]
\includegraphics[scale=0.50]{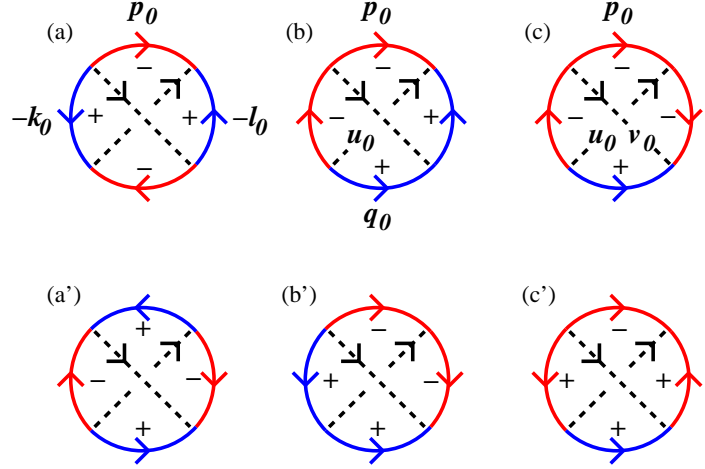}
\caption{%
    \label{fig:6terms}
    The six nonzero terms of Fig.\ \ref{fig:diags3}(a) arising from
    the decomposition (\ref{eq:Gexpand}) represented by
    Fig.\ \ref{fig:Gsplit}.
    The conventions for defining the direction of frequency flow
    are not given by the arrows
    here for the fermion lines but by the original diagram of
    Fig.\ \ref{fig:diags3}(a) for $p_0$, $q_0$, and $k_0$ and by
    the definition $l_0=-(p_0{+}q_0{+}k_0)$ for $l_0$.
    }
\end{figure}

In this new notation, Fig.\ \ref{fig:term}(a) shows an example of one
of the $2^{16}$ terms contained in the original diagram of
Fig.\ \ref{fig:diags3}(a).  It is easy to see that this term vanishes,
because there exists a loop, Fig.\ \ref{fig:term}(b), where all the
arrows have the same orientation.  If we do that loop integration first,
then we can close it in a half plane where there are no poles, and
we obtain zero.  There are only six terms that do not contain a similar
vanishing loop, and they are shown in Fig.\ \ref{fig:6terms}.
The terms represented by the bottom row are related to those of the
top row by the change of variables
\begin {equation}
   (P,Q,K) \to -(Q,P,K) ,
\label {eq:PQKsym}
\end {equation}
which is a symmetry of the original integrand
(\ref{eq:V3cross0}).%
\footnote{
  Another such symmetry is $K \to -(P{+}Q{+}K)$ with $P$ and $Q$
  unchanged.
}
This just represents a change of integration
variables, and so the contribution of the second row to the potential
will equal that of the first row.  We therefore need only evaluate
three terms, corresponding to Figs.\ \ref{fig:6terms}(a)--\ref{fig:6terms}(c).
Graphically, the operation (\ref{eq:PQKsym}) corresponds to
flipping the diagrams of Fig.\ \ref{fig:6terms} around the horizontal
axis and changing the designations $+ \leftrightarrow -$ on the
fermion lines.

The next step is to choose combinations of the three frequency
integration variables that make the integrals as simple as possible.
For each of the terms in Fig.\ \ref{fig:6terms}, one can find three
independent loops that have all arrows but one going around the
loop in the same direction.  Choose the frequency of the single
oppositely oriented line of each such loop to be an integration
variable.  In Fig.\ \ref{fig:6terms}, these three frequencies are shown
explicitly for each term in the first row, where
$l_0 \equiv -(p_0{+}q_0{+}k_0)$,
$u_0 \equiv -(q_0{+}k_0)$,
and
$v_0 \equiv p_0{+}k_0$.
By closing the frequency integration contours
in the appropriate half plane, one can pick up a single pole
for each corresponding to the labeled lines in the figure.
For instance, from Fig.\ \ref{fig:6terms}(a), one obtains the term
\begin {align}
  g^4 \phi_0^2 & \int_{\p\q\k}
    \frac{(E_\p - \eps_\p) (E_\q + \eps_\q)}
         {2E_\p \, 2E_\q \, 2E_\k \, 2E_{\p+\q+\k}}
    \int \frac{d p_0}{2\pi i} \> \frac{d q_0}{2\pi i} \> \frac{d l_0}{2\pi i}
    \>
    (p_0+E_\p-i\varepsilon)^{-1}
  \notag\\ & \qquad \times
     (-k_0-E_\k+i\varepsilon)^{-1}
     (-l_0-p_0-k_0+E_\q-i\varepsilon)^{-1}
     (-l_0-E_{\p+\q+\k}+i\varepsilon)^{-1}
  \notag\\ & \qquad \times
     (p_0+k_0-\half \eps_{\p+\k} + i\varepsilon)^{-1}
     (p_0+l_0-\half \eps_{\q+\k} + i\varepsilon)^{-1} ,
\\
\intertext{which integrates to}
  g^4 \phi_0^2 & \int_{\p\q\k}
    \frac{(E_\p - \eps_\p) (E_\q + \eps_\q)}
         {2E_\p \, 2E_\q \, 2E_\k \, 2E_{\p+\q+\k}} \>
    (E_\p + E_\q + E_\k + E_{\p+\q+\k})^{-1}
  \notag\\ & \qquad \times
    (E_\p + E_\k + \half \eps_{\p+\k})^{-1}
    (E_\p + E_{\p+\q+\k} + \half \eps_{\q+\k})^{-1} .
\end {align}
Doing all the terms of Fig.\ \ref{fig:6terms} similarly, we obtain the
following result for the contribution of the cross diagram to
the effective potential:
\begin {align}
  V_3^{(\otimes)} =
    g^4 \phi_0^2 \int_{\p\q\k}
    \Biggl\{ &
    \frac{ (E_\p-\eps_\p) }
         { 2 E_\p \, 2 E_\q \, 2 E_\k \, 2E_{\l} \, \S_{\p\l} }
    \left[
       \frac{(E_\q + \eps_\q)}{\S_{\p\k} \E_{\p\q\k\l}}
       - \frac{(E_\q - \eps_\q)}{\T_{\p\q,\k}}
         \left( \frac{1}{\S_{\p\k}} + \frac{1}{\S_{\q\k}} \right)
    \right]
\notag\\ &
   + (\p \leftrightarrow \q) \Biggr\} ,
\label {eq:V3cross}
\end {align}
where we introduce short-hand notation which will also be
convenient for other diagrams:
\begin {subequations}
\label {eq:shorthand}
\begin {align}
  \l &\equiv -(\p+\q+\k) ,
  \\
  \S_{\p\k} &\equiv E_\p + E_\k + \half \eps_{\p+\k} ,
  \\
  \T_{\p\q,\k} &\equiv E_\p + E_\q + \half \eps_{\p+\k} + \half \eps_{\q+\k} ,
  \\
  \E_{\p\q\k\l} &\equiv E_\p + E_\q + E_\k + E_\l .
\end {align}
\end {subequations}
The $(\p \leftrightarrow \q)$ term at the end of Eq.\ (\ref{eq:V3cross})
represents the second
row of Fig.\ \ref{fig:6terms}
[using the $(P,Q,K) \to -(Q,P,K)$ symmetry discussed earlier
combined with parity $(\p,\q,\k) \to -(\p,\q,\k)$].
One may drop the $(\p \leftrightarrow \q)$ in favor of multiplying
the rest of the expression by a factor of 2.


Some readers, used to perturbation theory about free Fermi gases,
may wonder at the absence of Fermi-sea step functions such as
$\theta(\mu - \eps_\p)$ in the expression (\ref{eq:V3cross}).  This is because
of the effects of the condensate $\phi_0$, which is large compared
to the chemical potential $\mu \sim \eps\phi_0$.  The system is
not describable as a small perturbation to a free Fermi sea.
In Appendix \ref{app:theta}, we give a brief, illustrative example of
what happens to a standard $\theta(\mu-\eps_\p)$ factor if one
adds a condensate $\phi_0$ and increases it to $\phi_0 \gg \mu$.



\subsubsection {The momentum integrals}
\label {sec:crossN}

The remaining integrals (\ref{eq:V3cross}) that we need to do can be
made dimensionless by rescaling momenta as
\begin {equation}
   \p \to (2 m \phi_0)^{1/2} \p ,
\label {eq:prescale}
\end {equation}
which has the effect of replacing $g^4 \phi_0^2$ by
$g^4 \phi_0^{-3} (2 m \phi_0)^{3d/2}$ outside the integral and
replacing $\eps_\p$ and $E_\p$ by the dimensionless versions
\begin {equation}
   \bar\eps_\p \equiv p^2 ,
   \qquad
   \bar E_\p \equiv (p^4+1)^{1/2}
\end {equation}
everywhere inside the integral.  Using the formula (\ref{eq:g2}) for
$g^2$, this can be written as
\begin {equation}
  V_3^{(\otimes)} =
    \phi_0 \left( \frac{m\phi_0}{2\pi} \right)^{d/2}
    \eps^2 \, {\cal K}_3^{(\otimes)}(\eps) ,
\end {equation}
where
\begin {equation}
  {\cal K}_3^{(\otimes)}(\eps) \equiv
    4 (4\pi)^{3d/2} \int_{\p\q\k}
    \frac{ 2 (\bar E_\p-\bar\eps_\p) }
         { 2 \bar E_\p \, 2 \bar E_\q \, 2 \bar E_\k \, 2\bar E_{\l}
           \, \bar\S_{\p\l} }
    \left[
       \frac{(\bar E_\q + \bar\eps_\q)}{\bar\S_{\p\k} \bar\E_{\p\q\k\l}}
       - \frac{(\bar E_\q - \bar\eps_\q)}{\bar\T_{\p\q,\k}}
         \left( \frac{1}{\bar\S_{\p\k}} + \frac{1}{\bar\S_{\q\k}} \right)
    \right]
\label {eq:K3cross}
\end {equation}
is a dimensionless function of $\eps$.
For a NNLO evaluation of the potential, we may evaluate
${\cal K}_3^{(\otimes)}$ at $\eps=0$, so that
\begin {equation}
  V_3^{(\otimes)} =
    \phi_0 \left( \frac{m\phi_0}{2\pi} \right)^{d/2}
    \eps^2 \, \bigl[ K_3^{(\otimes)} + O(\eps) \bigr] ,
\label {eq:V3crossK3}
\end {equation}
where the numerical constant $K_3^{(\otimes)}$ is the result of
the integral (\ref{eq:K3cross}) evaluated in exactly four dimensions.

For numerical evaluation, we use rotational invariance to rewrite
the momentum integrals as a six-dimensional integral over the magnitudes
of and angles between the momenta:
\begin {equation}
   \int \frac{d^4 p}{(2\pi)^4} \> \frac{d^4 q}{(2\pi)^4}
     \> \frac{d^4 k}{(2\pi)^4}
   = \frac{1}{(2\pi)^8} \int_0^\infty p^3 q^3 k^3 dp \> dq \> dk
     \int_0^\pi \sin^2\theta_\p \sin^2\theta_\q \sin \xi_\q
                 \> d\theta_\p \> d\theta_\q \> d\xi_\q ,
\end {equation}
corresponding to a choice of Cartesian coordinates aligned so that
\begin {subequations}
\begin {align}
   \k &= k \, (1,0,0,0) ,
\\
   \p &= p \, (\cos\theta_\p, \sin\theta_\p, 0, 0) ,
\\
   \q &= q \, (\cos\theta_\q, \sin\theta_\q \cos\xi_\q,
                           \sin\theta_\q \sin\xi_\q, 0) .
\end {align}
\end {subequations}
We then perform the integral (\ref{eq:K3cross})
numerically using adaptive Monte
Carlo integration.  The result is
\begin {equation}
    K_3^{(\otimes)} \simeq 0.15101 \,,
\label{eq:K3crossN}
\end {equation}
where $\simeq$ in this paper will mean that the result has an
estimated error of at most $\pm1$ in the last digit.
Equations (\ref{eq:V3crossK3}) and (\ref{eq:K3crossN}) represent our
final result for the cross diagram of Fig.\ \ref{fig:diags3}(a).


\subsection {Basic two-loop diagram through \boldmath$O(\eps^2)$}

The basic two-loop diagram of Fig.\ \ref{fig:diags2}(a) is given
by
\begin {equation}
  -i V_2^{(0)} =
  - (ig)^2 \int_{PQ} iG_{11}(P) \, iG_{22}(-Q) \, iD(P+Q) .
\end {equation}
Doing the frequency integrals,
\begin {equation}
  V_2^{(0)} =
  - g^2 \int_{\p\q}
    \frac{(E_\p-\eps_\p)(E_\q-\eps_\q)}
         {2 E_\p \, 2 E_\q \, \S_{\p\q}} \,.
\label {eq:V20}
\end {equation}
Rescaling momenta as in Eq.~(\ref{eq:prescale}),
\begin {equation}
  V_2^{(0)} =
    \phi_0 \left( \frac{m\phi_0}{2\pi} \right)^{d/2}
    \eps \, {\cal K}_2(\eps)
\end {equation}
with
\begin {equation}
  {\cal K}_2(\eps) =
    - 2(4\pi)^d \int_{\p\q}
    \frac{(\bar E_\p - \bar \eps_\p)(\bar E_\q - \bar \eps_\q)}
         {2 \bar E_\p \, 2 \bar E_\q \, \bar \S_{\p\q}} \,.
\end {equation}
If one sets $d{=}4$, one has ${\cal K}_2(0) = -C_2$ where
\begin {equation}\label{eq:C2-analytic}
   C_2 \equiv 
    2(4\pi)^4 \int \frac{d^4p}{(2\pi)^4} \> \frac{d^4q}{(2\pi)^4} \>
    \frac{(\bar E_\p - \bar \eps_\p)(\bar E_\q - \bar \eps_\q)}
         {2 \bar E_\p \, 2 \bar E_\q \, \bar \S_{\p\q}} \,,
\end {equation}
which gives
\begin {equation}
   C_2 \simeq 0.14424
\label {eq:C2N}
\end {equation}
upon numerical integration, all just as in Ref.\ \cite{Nishida&Son}.
For a NNLO calculation of the potential, however, we need the next
term in the expansion of ${\cal K}_2(\eps)$ in $\eps$.  We can obtain
this by rewriting the $d$ dimensional integration as
\begin {equation}
   \int \frac{d^d p}{(2\pi)^d} \> \frac{d^d q}{(2\pi)^d}
   = \frac{8}{(4\pi)^{d+\frac12} \Gamma(\frac{d}{2}) \, \Gamma(\frac{d-1}{2})}
     \int_0^\infty p^{d-1} q^{d-1} dp \> dq
     \int_0^\pi \sin^{d-2}\theta_{\p\q} \> d\theta_{\p\q} ,
\end {equation}
where $\theta_{\p\q}$ is the angle between $\p$ and $\q$, and then
expanding in $\eps$.  The result is
\begin {equation}
  {\cal K}_2(\eps) =
      - \bigl[ 1 + (\tfrac32 - \gamma - \ln2)\eps \bigr] C_2
      + \eps C_2^{(\rm log)}
      + O(\eps^2)
\end {equation}
where
\begin {subequations}
\label {eq:C2}
\begin {equation}
  C_2^{({\rm log})} =
    2(4\pi)^4 \int \frac{d^4p}{(2\pi)^4} \, \frac{d^4q}{(2\pi)^4} \>
    \frac{(\bar E_\p - \bar \eps_\p)(\bar E_\q - \bar \eps_\q)}
         {2 \bar E_\p \, 2 \bar E_\q \, \bar \S_{\p\q}}
    \, \ln\bigl( |\p| |\q| \sin\theta_{\p\q} \bigr) ,
\end {equation}
with numerical value
\begin {equation}
  C_2^{({\rm log})} \simeq 0.14238 \,.
\end{equation}
\end {subequations}


\subsection {Scalar loop with two self-energy insertions}
\label{sec:pipi}

The next diagrams we consider are those of Figs.\
\ref{fig:diags3}(f)--\ref{fig:diags3}(h),
which correspond to a scalar loop with two self-energy insertions
$\Pi$ and the corresponding counter-term diagrams.
Individually, the diagrams of Figs.\ \ref{fig:diags3}(f) and 
\ref{fig:diags3}(g) are
ultraviolet (UV) divergent; it is only in their combination that the
divergences are eliminated.  So we must be careful not to set
$d{=}4$ in integrations until we have organized the terms into
absolutely convergent integrals.  We will keep $d$ general
in what follows until near the end.

Together, Figs.\ \ref{fig:diags3}(f)--\ref{fig:diags3}(h) give
\begin {equation}
  -i V_3^{(\Pi\Pi)}
  =
  \half \int_{V} \bigl[i D(V)\bigr]^2
                 \bigl\{-i\bigl[\Pi(V)-\hat\Pi_0(V)\bigr]\bigr\}^2 ,
\label {eq:V3pipi1}
\end {equation}
where $V=(v_0,\v)$ is the frequency and momentum of the scalar line,
$\Pi$ is the one-loop scalar self-energy given by
\begin {equation}
  -i\Pi(V) = - (ig)^2 \int_P iG_{11}(P) \, i G_{22}(P-V) \,,
\label {eq:Pi1}
\end {equation}
and the corresponding counterterm $\hat\Pi_0$ is given by
Eq.~(\ref{eq:hatPi}).

The frequency integral in Eq.\ (\ref{eq:Pi1}) is straightforward,
especially using Eq.\ (\ref{eq:Gexpand}),
yielding
\begin {equation}
   \Pi(V) =
   g^2 \int_\p \frac{1}{2E_\p\,2E_{\p-\v}}\left[
      \frac{(E_\p+\eps_\p)(E_{\p-\v}+\eps_{\p-\v})}
           {v_0-E_\p-E_{\p-\v}+i\varepsilon}
     -\frac{(E_\p-\eps_\p)(E_{\p-\v}-\eps_{\p-\v})}
           {v_0+E_\p+E_{\p-\v}-i\varepsilon}
   \right] .
\label {eq:Pi}
\end {equation}
The momentum integral would be UV divergent if we set $d{=}4$.
We will isolate this divergence by isolating the large $\p$ behavior
of the integrand by writing
\begin {equation}
   \Pi(V) = \Pi_{\rm div}(V) + \Pi_{\rm reg}(V),
\label {eq:Pisplit}
\end {equation}
where the divergent piece of the momentum integral is
\begin {equation}
   \Pi_{\rm div}(V) \equiv
   g^2 \int_\p
      \frac{1}{v_0-\eps_\p-\eps_{\p-\v}+i\varepsilon}\,,
\label {eq:Pidiv1}
\end {equation}
and the remainder is
\begin {align}
   \Pi_{\rm reg}(V) =
   g^2 \int_\p & \Biggl\{
     \frac{1}{2E_\p\,2E_{\p-\v}}\left[
       \frac{(E_\p+\eps_\p)(E_{\p-\v}+\eps_{\p-\v})}
            {v_0-E_\p-E_{\p-\v}+i\varepsilon}
      -\frac{(E_\p-\eps_\p)(E_{\p-\v}-\eps_{\p-\v})}
            {v_0+E_\p+E_{\p-\v}-i\varepsilon}
     \right]
\notag\\ & \qquad
     - \frac{1}{v_0-\eps_\p-\eps_{\p-\v}+i\varepsilon}
   \Biggr\} .
\label {eq:Pireg}
\end {align}

The strategy here is to have chosen the form of $\Pi_{\rm div}$ to be simple
enough that we can manage to evaluate the integral in general dimension $d$.
The result of evaluating Eq.\ (\ref{eq:Pidiv1}) is%
\footnote{
  The dependence of (\ref{eq:PidivFormula}) only on the combination
  $-v_0 + \half\eps_\v$ is a result of Galilean invariance.
  Galilean invariance is broken by the condensate $\phi_0$, but
  $\phi_0$ does not appear in our definition of $\Pi_{\rm div}$.
}
\begin {align}
   \Pi_{\rm div}(V) &=
   - g^2 \,
   \Gamma\bigl(1-\tfrac{d}{2}\bigr)
   \left(\frac{m}{4\pi}\right)^{d/2}
   \bigl(-v_0 + \half \eps_\v - i\varepsilon\bigr)^{d/2-1}
\notag\\
   &=
   - \frac{\eps}{2} \, \Gamma\bigl(1-\tfrac{d}{2}\bigr) \,
   \bigl(-v_0 + \half \eps_\v\bigr)
   \left(\frac{-v_0 + \half \eps_\v - i\varepsilon}{2\phi_0}\right)^{-\eps/2}
   ,
\label {eq:PidivFormula}
\end {align}
where Eq.\ (\ref{eq:g2}) has been used for $g^2$.
Now use our split (\ref{eq:Pisplit}) of $\Pi$
to rewrite the contribution to the potential given by 
Eq.\ (\ref{eq:V3pipi1}) as
\begin {multline}
  V_3^{(\Pi\Pi)}
  =
  -\frac{1}{2i}
      \int_{V} \bigl[D(V)\bigr]^2 \biggl\{
         \bigl[\Pi_{\rm reg}(V)\bigr]^2
         + 2 \Pi_{\rm reg}(V) \bigl[\Pi_{\rm div}(V) - \hat\Pi_0(V)\bigr]
\\
         + \bigl[\Pi_{\rm div}(V) - \hat\Pi_0(V)\bigr]^2
      \biggr\}
   .
\label {eq:V3pipi2}
\end {multline}
In dimensional regularization, the terms which do not involve
$\Pi_{\rm reg}$ integrate to zero because of their scaling
properties.  For example, changing integration variables
$(v_0,\v) \to (\lambda^2 v_0, \lambda \v)$ for an arbitrary
constant $\lambda$,
\begin {equation}
   \int \frac{dv_0}{2\pi} \, \frac{d^dv}{(2\pi)^d}
      \bigl[D(V)\bigr]^2 \bigl[\Pi_{\rm div}(V)\bigr]^2
   =
   \int \frac{\lambda^2 dv_0}{2\pi} \, \frac{\lambda^d d^dv}{(2\pi)^d}
      \bigl[\lambda^{-2} D(V)\bigr]^2
      \bigl[\lambda^{d-2}\Pi_{\rm div}(V)\bigr]^2
\end {equation}
So this integral must equal itself times $\lambda^{3(d-2)}$ and so must
vanish.
For similar reasons, we may dispense which each term generated by
expanding the square in
\begin {equation}
  -\frac{1}{2i}
      \int_{V} \bigl[D(V)\bigr]^2
         \bigl[\Pi_{\rm div}(V) - \hat\Pi_0(V)\bigr]^2
  = 0 .
\end {equation}

For the first term in Eq.\ (\ref{eq:V3pipi2}), we continue by performing
the $v_0$ integration, using Eq.\ (\ref{eq:Pireg}) for $\Pi_{\rm reg}$.
One may avoid the bother of dealing with the double pole from $D(K)$
by closing in the upper-half plane.  The result is
\begin {multline}
  -\frac{1}{2i}
      \int_{V} \bigl[D(V)\bigr]^2 \bigl[\Pi_{\rm reg}(V)\bigr]^2
\\ =
    - \frac{g^4}{2} \int_{\p\q\k} \Biggl\{ 
    \frac{ (E_\p-\eps_\p)(E_\k-\eps_\k) }
         { 2 E_\p \, 2 E_\q \, 2 E_\k \, 2 E_\l \, \S_{\p\k}^2}
       \biggl[
         \frac{(E_\q-\eps_q)(E_\l-\eps_\l)}{\S_{\q\l}}
         + \frac{(E_\q+\eps_q)(E_\l+\eps_\l)}{\E_{\p\q\k\l}}
\\
         - \frac{2 E_\q \, 2E_\l}{(E_\p+\eps_\q+E_\k+\eps_{\l})}
       \biggl]
     + (\p\k \leftrightarrow \q\l)
     \Biggr\} ,
\end {multline}
where we have used the notation of Eqs.\ (\ref{eq:shorthand}) and the
momentum naming conventions shown in Fig.\ \ref{fig:diags3}(f)
(so $\v = \p+\k$).
This integral is absolutely convergent in $d{=}4$, and so we may
evaluate it numerically just as we did for the cross diagram in
Sec.\ \ref{sec:crossN}, giving
\begin {equation}
  -\frac{1}{2i}
      \int_{V} \bigl[D(V)\bigr]^2 \bigl[\Pi_{\rm reg}(V)\bigr]^2
  =
    \phi_0 \left( \frac{m\phi_0}{2\pi} \right)^{d/2}
    \eps^2 \, \bigl[ K_3^{(\Pi\Pi,1)} + O(\eps) \bigr]
\label {eq:sigsig1}
\end {equation}
with
\begin {equation}
    K_3^{(\Pi\Pi,1)} \simeq 0.006753 \,.
\label {eq:K3pipiN}
\end {equation}

Finally, we need the second term from the right-hand side of 
Eq.\ (\ref{eq:V3pipi2}).
Taking $\Pi_{\rm reg}$ and $\Pi_{\rm div}$ from Eqs.\ (\ref{eq:Pireg}) and
(\ref{eq:PidivFormula}), and closing the $v_0$ integration in the
upper half plane to avoid the branch cut in $\Pi_{\rm div}$,
one finds
\begin {align}
  -\frac{1}{i}
      \int_{V} \bigl[D(V)\bigr]^2
      &\Pi_{\rm reg}(V)\bigl[\Pi_{\rm div}(V) - \hat\Pi_0(V) \bigr]
\notag\\
    &= g^2 \int_{\p\k}
    \frac{ (E_\p-\eps_\p)(E_\k-\eps_\k) }
         { 2 E_\p \, 2 E_\k \, \S_{\p\k}^2}
    \bigl[\Pi_{\rm div}(V) - \hat\Pi_0(V) \bigr]
    \Big|_{-v_0+\half\eps_\v = S_{\p\k}}
\notag\\
    &= g^2 \int_{\p\k}
    \frac{ (E_\p-\eps_\p)(E_\k-\eps_\k) }
         { 2 E_\p \, 2 E_\k \, \S_{\p\k}}
       \left[
           - \frac{\eps}{2} \, \Gamma\bigl(1{-}\tfrac{d}{2}\bigr)
           \left(\frac{S_{\p\k}}{2\phi_0}\right)^{-\eps/2}
           - 1
       \right] .
\end {align}
If we expand about four dimensions, we obtain absolutely convergent
integrals at every order in $\eps$.  We are therefore free to expand
the integrand in $\eps$ to obtain
\begin {equation}
    [1+O(\eps)] \, \frac{\eps g^2}{2} \int_{\p\k}
    \frac{ (E_\p-\eps_\p)(E_\k-\eps_\k) }
         { 2 E_\p \, 2 E_\k \, \S_{\p\k}}
       \left[
         1 - \gamma - \ln\left(\frac{S_{\p\k}}{2\phi_0}\right)
       \right] .
\end {equation}
Rescaling momenta in the usual way then gives
\begin {multline}
  -\frac{1}{i}
      \int_{V} \bigl[D(V)\bigr]^2
      \Pi_{\rm reg}(V)\bigl[\Pi_{\rm div}(V) - \hat\Pi_0(V) \bigr]
\\ =
    \phi_0 \left( \frac{m\phi_0}{2\pi} \right)^{d/2}
    \Bigl\{
      \bigl[ \half(1-\gamma+\ln2) C_2 - C_2^{({\rm log,\S})} \bigr] \eps^2
      + O(\eps^3)
    \Bigr\}
\label {eq:sigsig2}
\end {multline}
with $C_2$ given by Eqs.\ (\ref{eq:C2-analytic}) and (\ref{eq:C2N}) and
\begin {subequations}
\label {eq:C2logS}
\begin {equation}
  C_2^{({\rm log,\S})} \equiv
    2(4\pi)^4 \int \frac{d^4p}{(2\pi)^4} \, \frac{d^4q}{(2\pi)^4} \>
    \frac{(\bar E_\p - \bar \eps_\p)(\bar E_\q - \bar \eps_\q)}
         {2 \bar E_\p \, 2 \bar E_\q \, \bar \S_{\p\q}}
    \, \half \ln(\bar\S_{\p\q})
\end {equation}
giving
\begin {equation}
  C_2^{({\rm log,\S})} \simeq 0.19408 \,.
\end {equation}
\end {subequations}
Our final result for the contribution of Figs.\ 
\ref{fig:diags3}(f)--\ref{fig:diags3}(h) to
the effective potential is the sum of Eqs.\ (\ref{eq:sigsig1}) and
(\ref{eq:sigsig2}):
\begin {equation}
  V_3^{(\Pi\Pi)} =
    \phi_0 \left( \frac{m\phi_0}{2\pi} \right)^{d/2}
    \Bigl\{
      \bigl[ K_3^{(\Pi\Pi,1)} +
             \half(1-\gamma+\ln2) C_2 - C_2^{({\rm log,\S})} \bigr] \eps^2
      + O(\eps^3)
    \Bigr\} .
\label {eq:V3pipi}
\end {equation}


\subsection {Remaining diagrams}
\label {sec:remaining}

The remaining diagrams are relatively easy, and we summarize results
in Appendix \ref{app:summary}.  Here, we will just make a few
comments on method.

A simple way to handle diagrams with self-energy insertions is to
consider the diagram without any such insertions, and then replace
$\eps_\p$ by
$\eps_\p^\mu \equiv \eps_\p-\mu$
in both $\eps_\p$ and
$E_\p = (\eps_\p^2+\phi_0^2)^{1/2}$ for fermion energies, and
replace $\half\eps_\v$ by $\half\eps_\v - 2\mu$ for boson
energies.  For instance, to simultaneously evaluate all of
the two-loop diagrams of Fig.\ \ref{fig:diags2}(b)--\ref{fig:diags2}(d) with
chemical potential insertions, replace Eq.\ (\ref{eq:V20}) for
the basic two-loop result $V_2^{(0)}$ by
\begin {equation}
  - g^2 \int_{\p\q}
    \frac{(E_\p^\mu-\eps_\p^\mu)(E_\q^\mu-\eps_\q^\mu)}
         {2 E_\p^\mu \, 2 E_\q^\mu \,
         (E_\p^\mu + E_\q^\mu + \half\eps_{\p+\q} - 2\mu)} \,,
\end {equation}
where $E_\p^\mu \equiv [(\eps_p-\mu)^2+\phi_0^2]^{1/2}$.
Then Taylor expand the integrand to the desired order in $\mu$,
which in this case is first order.
The same method can be used on the one-loop integrals
of Fig.\ \ref{fig:diags1} starting from the basic one-loop
integral of Fig.\ \ref{fig:diags1}(a),%
\footnote{
  There are similar issues as footnote \ref{foot:Cinf} concerning
  the frequency integral and the behavior of its integrand at infinity.
  In general regularization schemes, this could be avoided by defining
  the integrals here with $\phi_0$
  (and $\mu$) independent subtractions, which will not affect the
  determination of $\xi$ from the effective potential:
  $   i \int_P \bigl\{ \ln \det \bigl[ -i G^{-1}(P;\phi_0) \bigr]
                     - \ln \det \bigl[ -i G^{-1}(P;0) \bigr] \bigr\}
    = - \int_\p (E_\p - \eps_\p) $.
  But, in dimensional regularization, the subtracted term vanishes
  anyway
  by scaling arguments similar to those reviewed in Sec.\ \ref{sec:pipi}.
}
\begin {equation}
   V_1^{(0)} = i \int_P \ln \det \bigl[ -i G^{-1}(P) \bigr]
   = - \int_\p E_\p \,.
\label {eq:V0}
\end {equation}


\section {Result for \boldmath$\xi$}
\label {sec:xi}

Combining the results given previously and in Appendix
\ref{app:summary}, the full effective potential at NNLO in $\eps$ is
\begin {equation}\label{Vcombined}
   V(\phi_0) =
   \left( \frac{m\phi_0}{2\pi}\right)^{d/2}
   \left[
     \frac{\phi_0}{3} (1 + a_1 \eps + a_2 \eps^2)
     - \frac{\mu}{\eps} (1 + b_1 \eps + b_2 \eps^2)
     - \frac{\mu^2}{2\phi_0}
  \right]
  + O(\eps^3) ,
\end {equation}
where $\phi_0$ is treated as $O(1)$ and $\mu$ as $O(\eps)$.
The various numerical coefficients are
\begin {align}
  a_1 &= \half \bigl( \tfrac73 - \gamma - \ln2 \bigr) - 3 C_2
       \simeq 0.09877 ,
\\
  a_2 &=
  \tfrac18 \bigl( \tfrac73 - \gamma - \ln2 \bigr)^2
  + \tfrac{19}{72}
  + 3 \Bigl[
    \bigr(-1+\half\gamma+\tfrac32 \ln 2\bigr) C_2
    + C_2^{(\rm log)}
    - C_2^{({\rm log},\S)}
    + K_3
    \Bigr]
\notag\\ & \qquad
  \simeq -0.15840 \, ,
\\
  b_1 &= \half \bigl( \half - \gamma + \ln 2 \bigr)
       \simeq 0.30797 \,,
\\
  b_2 &=
  \tfrac18 \bigl( \half - \gamma + \ln2 \bigr)^2
  + \tfrac1{32}
  - K_2^{(\mu)}
  \simeq 0.33703 \,,
\end {align}
where
\begin {equation}
   K_3 \equiv
      K_3^{(\otimes)}
    + K_3^{(\Pi\Pi,1)}
    + K_3^{(\Sigma\Sigma)}
    + K_3^{(\widetilde\Pi\widetilde\Pi)}
  \simeq -0.18348 \,.
\end {equation}
The ratio $\xi$ can be computed from $V(\phi_0,\mu)$ by the
procedure used in Ref.\ \cite{Nishida&Son}.
First, we determine the expectation $\phi_0$ which minimizes the
potential:
\begin {equation}
  \phi_0 =
  \frac{2\mu}{\eps} \bigl[
    1 + 0.12586\,\eps + 0.56845 \, \eps^2 + O(\eps^3)
  \bigr] .
\label {eq:phi0}
\end {equation}
Then we determine the fermion number density $n$ from
the pressure $P = -V(\phi_0)$, giving
\begin {equation}
  n = \frac{\partial P}{\partial\mu}
  = - \frac{\partial V}{\partial\mu}
  = \frac{1}{\eps} \left( \frac{m\phi_0}{2\pi}\right)^{d/2}
    \left[
      1 + b_1 \eps + b_2 \eps^2
      + \frac{\eps\mu}{\phi_0}
      + O(\eps^3)
    \right] .
\label {eq:n}
\end {equation}
Next, we take the formula for the Fermi energy $\eps_{\rm F}(n)$ of
a $d$-dimensional ideal Fermi gas with the same density,
\begin {equation}
  \eps_{\rm F} =
  \frac{2\pi}{m} \left[
     \half \, \Gamma \bigl( \tfrac{d}{2}+1 \bigr) \, n \right]^{2/d} .
\end {equation}
For the density (\ref{eq:n}), this then gives
\begin {equation}
  \xi \equiv \frac{\mu}{\eps_{\rm F}} =
  \frac{\eps^{2/d}\mu}{\phi_0} \left\{
      \half \, \Gamma \bigl( \tfrac{d}{2}+1 \bigr)
    \left[
      1 + b_1 \eps + b_2 \eps^2
      + \frac{\eps\mu}{\phi_0}
      + O(\eps^3)
    \right]
  \right\}^{-2/d} .
\end {equation}
Substituting in the expectation (\ref{eq:phi0}) for $\phi_0$,
and rewriting $\eps^{2/d} = \eps^{\eps/2d} \eps^{1/2}$,
produces the final result (\ref{eq:xi}) through NNLO.
The additional logarithmic term shown in Eq.\ (\ref{eq:xi}) at the
order beyond NNLO is the subject of the next section.


\section {The \boldmath$\eps$ expansion beyond next-to-next-to-leading order}
\label {sec:beyond NNLO}

\subsection {General}

In this section, we discuss a difficulty that arises in applying the
diagrammatic expansion of Sec.\ \ref{sec:review} if one were to proceed
to yet one higher order in $\eps$, attempting to evaluate the effective
potential through $O(\eps^3)$ [and so $\xi$ through $O(\eps^{9/2})$].
The problem is an infrared problem arising from the fact that the
scalar excitations, unlike the fermionic excitations, are not gapped.

\begin{figure}[t]
\includegraphics[scale=0.50]{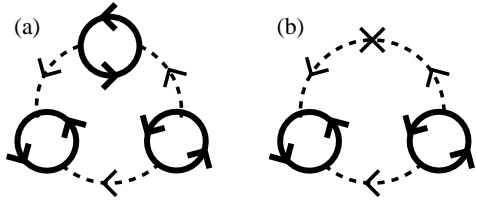}
\caption{%
    \label{fig:IRdiag}
    Two examples of diagrams that produce logarithmic infrared divergences.
    }
\end{figure}

\begin{figure}[t]
\includegraphics[scale=0.50]{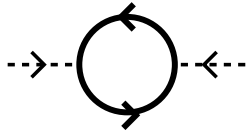}
\caption{%
    \label{fig:PiTilde}
    The self-energy $\widetilde\Pi$ that mixes $\varphi$ with $\varphi^*$.
    }
\end{figure}

For the sake of specificity, consider the $O(\eps^3)$ contribution to
the effective potential made by the diagram of Fig.\ \ref{fig:IRdiag}(a).
Together with the corresponding counterterm diagram, this
gives a contribution to the effective potential
\begin {equation}
   V_4^{(\rm example)} =
   - \frac{1}{i} \int_V
   \bigl[ D(V) \bigr]^2 D(-V)
   \bigl[ \Pi(V) - \hat\Pi_0(V) \bigr] \bigl[ \widetilde\Pi(V) \bigr]^2 ,
\end {equation}
where the one-loop self-energy $\Pi$ is given by
Fig.\ \ref{fig:scalar} and $\widetilde\Pi$ by Fig.\ \ref{fig:PiTilde}.
Let us now explore the contribution to this integral from small
scalar frequency and momentum: $|v_0| \ll \phi_0$ and
$\half\eps_\v \ll \phi_0$.
At small $V$, we approximate $V$ as zero inside the self-energies,
and this region of integration contributes
\begin {equation}
   V_4^{(\text{ex, small $V$})} \sim
   - \Pi(0) \bigl[ \widetilde\Pi(0) \bigr]^2
   \frac{1}{i}
   \int_V
   \bigl[ D(V) \bigr]^2 D(-V) ,
\end {equation}
where the integration is restricted to small $V$.
[Note that $\hat\Pi_0(0)=0$.]
The frequency integral is simple, giving
\begin {equation}
   V_4^{(\text{ex, small $V$})} \sim
   \Pi(0) \bigl[ \widetilde\Pi(0) \bigr]^2
   \int_\v
   \frac{1}{\eps_\v^2} \,.
\label {eq:IRdiv}
\end {equation}
The momentum integral is IR divergent in $d \le 4$.
So our evaluation of diagrams has broken down in the infrared
if $\Pi(0)$ and $\widetilde\Pi(0)$ are nonzero.%
\footnote{
  If one blindly tried to regulate the IR divergences using dimensional
  regularization (which we have previously used only to regulate the UV),
  then the IR momentum integral in Eq.\ (\ref{eq:IRdiv}) would generate a
  factor of $1/\eps$, which would in any case destroy the $\eps$
  counting of Sec.\ \ref{sec:review}.  The integral does not generate
  zero (by scaling arguments like those of Sec.\ \ref{sec:pipi}) because
  the $1/\eps_\v^2$ integrand is an approximation valid only for
  $|\v| \ll \phi_0$.  The behavior of the full result changes for
  $|\v| \gtrsim \phi_0$, and so $\phi_0$ provides a scale.
  (We say ``blindly tried to regulate'' because dimensional
  regularization throws away nonlogarithmic divergences, and it
  does not distinguish between IR and UV logarithmic divergences.
  This means
  it can be a dangerous procedure unless you already know that all
  divergences will cancel in the final result.)
}

Another example of a diagram producing similar problems is shown
in Fig.\ \ref{fig:IRdiag}(b).
Adding one or more additional self-energy or chemical potential insertions to
the scalar loop in either diagram would generate even more severe (power-law)
infrared divergences.

We shall see in a moment that $\Pi(0)$ and $\widetilde\Pi(0)$ are both
$O(\mu)$.  The problem, then, is that when $\v$ is small enough
that $\eps_\v$ is of order $\mu \ll \phi_0$, it is no longer a
good approximation to treat the scalar self-energies, or the chemical
potential $2\mu$, as perturbations to $D^{-1} = -v_0 + \half\eps_\v$.
One must therefore resum $\Pi(0)$, $\widetilde\Pi(0)$, and $2\mu$ into
scalar propagators in order to recover a well-behaved perturbation
theory.  Equivalently, one must use the appropriate low-energy
scalar effective
theory when evaluating the contribution of low-energy scalars
($|v_0| \ll \phi_0$) to the effective potential.
At low energy, the effective scalar propagator would be generated
by the effective interactions
\begin {align}
   {\cal L}_{\rm eff} &\simeq
   \varphi^* \left( i \partial_t + \frac{\nabla^2}{4m} \right) \varphi
     + 2\mu \varphi^* \varphi
     - \Pi(0) \, \varphi^* \varphi
     - \widetilde\Pi(0) \, \half(\varphi \varphi + \varphi^* \varphi^*)
\notag\\
   &\equiv
   \half
   \begin{pmatrix} \varphi \\ \varphi^* \end{pmatrix}^\dagger
   {\cal D}^{-1}
   \begin{pmatrix} \varphi \\ \varphi^* \end{pmatrix}
\label {eq:Leff}
\end {align}
with corresponding propagator
\begin {equation}
   {\cal D}(P) =
   \begin {pmatrix}
      p_0 - \half\eps_\p + 2\mu - \Pi(0) + i\varepsilon
         && - \widetilde\Pi(0) \\
      - \widetilde\Pi(0)
         && - p_0 - \half\eps_\p + 2\mu - \Pi(0) + i\varepsilon
   \end {pmatrix}^{-1} .
\end {equation}

To better understand the structure of this propagator, we now
evaluate $\Pi(0)$ and $\tilde\Pi(0)$.  From Eq.\ (\ref{eq:Pi}),
\begin {equation}
   \Pi(0) =
   - g^2 \int_\p \frac{E_\p^2+\eps_\p^2}{4E_\p^3} .
\end {equation}
$\widetilde\Pi(0)$ is given by
\begin {equation}
  \widetilde\Pi(0) =
  \frac{g^2}{i} \int_P \bigl[ G_{12}(P) \bigr]^2 = 
  g^2 \int_\p \frac{\phi_0^2}{4E_\p^3} .
\end {equation}
In dimensional regularization, the momentum integrals give
\begin {equation}
  \Pi(0) =
  - \frac{g^2}{\phi_0}
  \left( \frac{m \phi_0}{4\pi}\right)^{d/2}
  \frac{\bigl(1+\frac{d}{2}\bigr) \, \Gamma\bigl(\half-\frac{d}{4}\bigr)}
       {4 \, \Gamma\bigl(\half+\frac{d}{4} \bigr)}
  = 3\mu \, [1 + O(\eps)]
\end {equation}
and
\begin {equation}
  \widetilde\Pi(0) =
  \frac{(d-2)}{(d+2)} \, \Pi(0)
  = \mu \, [1 + O(\eps)] ,
\end {equation}
where we have used Eqs.\ (\ref{eq:g2}) and (\ref{eq:phi0}) for
$g^2$ and $\phi_0$.
To leading order in $\eps$ for scalars with energy of order $\mu$,
the effective low-energy interactions (\ref{eq:Leff}) are then
\begin {equation}
   {\cal L}_{\rm eff} \simeq
   \varphi^* \left( i \partial_t + \frac{\nabla^2}{4m} \right) \varphi
     - \half \mu (\varphi + \varphi^*)^2
\end {equation}
with
\begin {equation}
   {\cal D}(P) =
   \begin {pmatrix}
      p_0 - \half\eps_\p - \mu + i\varepsilon
         && - \mu \\
      - \mu
         && - p_0 - \half\eps_\p - \mu + i\varepsilon
   \end {pmatrix}^{-1} .
\end {equation}
The imaginary part of $\varphi$ remains gapless, which
must occur in any consistent approximation since it corresponds to a
Goldstone boson.

With this formula, we can now compute the leading contribution to the
effective potential from low-energy scalars and see that all is well.
Analogous to the fermionic result (\ref{eq:V0}), it is
\begin {align}
  V^{(\text{soft $\varphi$})} &\simeq
   - \frac{i}{2} \int_P \ln\det\bigl[-i{\cal D}^{-1}(P)\bigr]
  = - \frac{i}{2} \int_P
     \ln\Bigl[p_0^2 - \half\eps_\p (\half \eps_\p + 2\mu) + i\varepsilon\Bigr]
\notag\\
  &= \half \int_\p \Bigl[ \half\eps_\p (\half \eps_\p + 2\mu) \Bigr]^{1/2} .
\label {eq:Vsoft}
\end {align}
This integral has no infrared divergences.
If we naively expanded
the integrand
in powers of $\mu$, we would obtain infrared divergences starting
at third order $\sim \mu^3 \int_\p \eps_\p^{-2}$.  This is the same
order as the result (\ref{eq:IRdiv}) for Fig.\ \ref{fig:IRdiag}
that started this discussion.  If we instead integrate
Eq.\ (\ref{eq:Vsoft}) up to a UV momentum cutoff $(4m\Lambda)^{1/2}$ for
the effective theory, we find an expansion in $\mu$ of the form
\begin {equation}
  V^{(\text{soft $\varphi$})} \simeq
  m^2 \left[ \#\Lambda^3 + \#\mu\Lambda^2 + \#\mu^2\Lambda +
             \# \mu^3 \ln\left(\frac{\mu}{\Lambda}\right) + \#\mu^3
             + \cdots \right] .
\label {eq:Vsoftform}
\end {equation}
The soft effective scalar theory breaks down at $\eps_\p \gtrsim \phi_0$,
so one should very roughly think of the energy scale $\Lambda$ as
of order $\phi_0$.

The moral of this story is that one will have to make a proper treatment
of low-energy scalar fields in order to go to higher orders in $\eps$
than the NNLO calculation performed in the bulk of this paper.  In particular,
the approach of Sec.\ \ref{sec:review} would lead to logarithmic
divergences in the effective potential at $O(\eps^3)$ and worse
divergences at higher order, but a proper
resummation of soft scalar physics will resolve these divergences into
$O\bigl(\mu^3 \ln(\mu/\phi_0)\bigr) = O(\eps^3 \ln\eps)$.
This corresponds to a correction to $\xi$ of order
$\eps^{\eps/2d} \eps^{9/2} \ln\eps$.


\subsection{The coefficient of the logarithm}

Though we are not prepared to make a full calculation of the order
beyond NNLO in the $\eps$ expansion, it is simple to extract the
coefficient of the logarithm at that order from the preceding
discussion.  Naively expand the integrand in the formula
(\ref{eq:Vsoft})
for the soft contribution $V^{(\text{soft $\varphi$})}$
to third order in $\mu$.  The third-order term is
\begin {equation}
  V^{(\mu^3)} =
  \mu^3 \int_\p \frac{1}{\eps_\p^2} .
\end {equation}
In four dimensions, this is the logarithmically divergent integral
\begin {equation}
  V^{(\mu^3)} =
  \frac{\mu^3}{8\pi^2} \int_0^\infty \frac{p^3 \> dp}{\eps_\p^2}
  =
  \frac{m^2 \mu^3}{4\pi^2} \int_0^\infty \frac{d\eps_\p}{\eps_\p} \,.
\end{equation}
But, from the discussion surrounding Eq.\ (\ref{eq:Vsoftform}), 
we know that this
logarithm is cut off in the infrared by the energy scale $\mu$, due to
the necessity of resummation, and in
the ultraviolet by the energy scale $\phi_0$, the energy scale where
the self-energies are no longer well approximated by their zero-momentum
values.
Thus, even though we cannot easily
compute the constant under the logarithm, we can write%
\footnote{
  The coefficient of this logarithm has also been computed by
  Y.~Nishida \cite{NishidaLog}.}

\begin {equation}
  V^{(\mu^3)} =
  \frac{m^2 \mu^3}{4\pi^2} \left[ \ln\left( \frac{\phi_0}{\mu} \right)
                                  + O(1) \right] .
\label {eq:Vmu3log}
\end{equation}

We can now use this to obtain the explicit
logarithm shown in our final result (\ref{eq:xi}) for $\xi$
beyond NNLO by including Eq.\ (\ref{eq:Vmu3log}) in the analysis of
Sec.\ \ref{sec:xi}.
Because $\partial V^{(\mu^3)}/\partial\phi_0$ does not have a logarithm,
there is no logarithm in the NNNLO result for the location $\phi_0$ of
the minimum of the effective potential.  The logarithm appears
in $\xi$ only through its effect on the density
$n = - \partial V/\partial \mu$.
Equation (\ref{eq:n}) for $n$ is modified to
\begin {equation}
  n =
 \frac{1}{\eps} \left( \frac{m\phi_0}{2\pi}\right)^{d/2}
    \left[
      1 + b_1 \eps + b_2 \eps^2
      + \frac{\eps\mu}{\phi_0}
      - \frac{3\eps\mu^2}{\phi_0^2} \ln \frac{\phi_0}{\mu}
      + O(\eps^3)
    \right] ,
\end {equation}
giving
\begin {equation}
  \xi \equiv \frac{\mu}{\eps_{\rm F}} =
  \frac{\eps^{2/d}\mu}{\phi_0} \left\{
      \half \, \Gamma \bigl( \tfrac{d}{2}+1 \bigr)
    \left[
      1 + b_1 \eps + b_2 \eps^2
      + \frac{\eps\mu}{\phi_0}
      - \frac{3\eps\mu^2}{\phi_0^2} \ln \frac{\phi_0}{\mu}
      + O(\eps^3)
    \right]
  \right\}^{-2/d} .
\end {equation}
Substituting the expectation (\ref{eq:phi0}) for $\phi_0$ then
produces the NNNLO logarithm shown in the final result (\ref{eq:xi})
for $\xi$.


\section {Extrapolation of \boldmath$\xi$ at \boldmath$d{=}3$}
\label {sec:extrapolate}

Because of the large relative size
of the $O(\eps^{7/2})$ term in our result (\ref{eq:xi}) for $\xi$,
the next-to-next-to-leading order (NNLO) result will likely only be useful
in conjunction with more sophisticated analysis of dimension dependence
than the naive prescription of setting $\eps{=}1$.
In this respect, the situation appears somewhat analogous to the
$\epsilon$ expansion for the critical exponent $\omega$ in the Ising
model.  $\omega$ is the exponent characterizing corrections to scaling,
and its $\epsilon$ expansion is \cite{current exponents}
\begin {equation}
  \omega_{\rm Ising} =
  \eps - 0.62963 \, \eps^2 + 1.61822 \, \eps^3 + O(\eps^4) .
\end {equation}

Despite the large size of the NNLO $\eps^3$ term, historical analysis
using the terms shown above gave $\omega \simeq 0.79$
with a simple Borel-Pad\'e approximation \cite{BGZ2}.
The latter is within a few percent of the correct result.%
\footnote{
  See, for instance, the results for $\theta = \omega\nu$ and $\nu$
  from three-dimensional (3D) series techniques, the $\eps$
  expansion, numerical Monte Carlo simulations, and experiments all
  reviewed in Ref.\ \cite{current exponents}
  for the $O(N)$ model.  (The Ising universality class corresponds to
  $N=1$ in these tables.)  The current results presented there for
  $\omega$ from 3D series techniques and the $\eps$ expansion
  are 0.799(11) and 0.814(18).
}

Unfortunately, the ratio $\xi$ does not have as simple an analytic
structure as do critical exponents.  Our result for $\xi$ may be
written as
\begin {equation}
   \xi
   = \half \eps^{1+2/d} \, b(\eps)
\end {equation}
with
\begin {equation}
   b(\eps) = 1
        - 0.04916 \eps
        - 0.95961 \, \eps^2
        - \tfrac38 \, \eps^3 \ln\eps
        + O(\eps^{9/2})
\label {eq:b}
\end {equation}
Critical exponents have a simple
asymptotic expansion in powers of $\eps$.
But, even if we factor out the overall $\eps^{1+2/d}$ and focus
only on $b(\eps)$ above, our expansion contains powers of $\ln\eps$.

The Borel transform of a series
\begin {equation}
   f(\eps) = \sum_n f_n \eps^n
\end {equation}
is the faster-converging series
\begin {equation}
   F(t) = \sum_n \frac{f_n}{n!} \, t^n .
\end {equation}
The original $f(\eps)$ may be recovered from its Borel transform by
\begin {equation}
   f(\eps) = \int_0^\infty dt \> e^{-t} F(\eps t)
\label {eq:Ftof}
\end {equation}
if $F(t)$ does not have any singularities on the positive real axis
which make the integral ill defined.  The standard approach for
critical exponents is to fit some type of Pad\'e-like approximation
to the Borel transform.  A simple $[M/N]$ Pad\'e approximation would be
\begin {equation}
   F(t) = \frac{1 + p_1 t + p_2 t^2 + \cdots p_M t^M}
               {1 + q_1 t + q_2 t^2 + \cdots q_N t^N} \,,
\label {eq:Pade}
\end {equation}
where we assume $f(\eps)$ is normalized so that $f(0)=1$.
More sophisticated versions have been used for accurate
estimates to critical exponents from high-order $\epsilon$ expansions.
However, such estimates must necessarily break down for the Borel
transform $B(t)$ of (\ref{eq:b}) because of the $\ln\eps$ terms.
The appearance of $\ln\eps$ in the small $\eps$ expansion of
$b(\eps)$ gives rise to $\ln t$ terms in the small $t$ expansion
of its Borel transform, and these are not accommodated by
Pad\'e approximants such as (\ref{eq:Pade}).%
\footnote{
   Specifically, the Borel transform of
   $b(\eps) = 1 + \alpha \eps + \beta \eps^2 + \gamma \eps^3 \ln\eps
                + \delta \eps^3$
   is
   $B(t) = 1 + \alpha t + \frac{1}{2!} \beta t^2
             + \frac{1}{3!} \gamma t^3 \ln t
             + \frac{1}{3!} [\delta - \gamma \, \psi(4)] t^3$,
   where $\psi(z)$ is the digamma function.
   This can be demonstrated by checking Eq.\ (\ref{eq:Ftof}).
}

Clearly, what is needed to make full use of $\eps$ expansion results
is a full understanding of the analytic
structure of $\xi(\eps)$ in $\eps$, in order to inform the strategy for
how best to extrapolate to $d{=}3$.  Nonetheless, it is interesting to
see what happens if one naively extrapolates $\xi(\eps)$ using
simple Pad\'e estimates (\ref{eq:Pade}) to the Borel transform
$B(t)$.  This was carried out by Nishida and Son in Ref.\ \cite{Nishida&Son2},
constraining the Pad\'e approximants by (i) the next-to-leading-order 
(NLO) result near the expansion of $\xi$ in $4{-}d$, and
(ii) a next-to-leading-order result for the expansion of $\xi$ in
$d-2$.  Since the leading-order $4{-}d$ result has already been
used to normalize $b(0)=1$ above, this represents three constraints
on $B(t)$, and so the possible Pad\'e approximants are those with
$M+N=3$.  Extrapolating to three dimensions, they then found
$\xi = 0.391$, 0.364, and 0.378 with $[3/0]$, $[1/2]$, and $[0/3]$
Pad\'e approximants.  They did not find any solution satisfying the
constraints for a $[2/1]$ approximant.  These values for $\xi$ span
$0.378\pm0.014$.

\begin{figure}[t]
\includegraphics[scale=0.50]{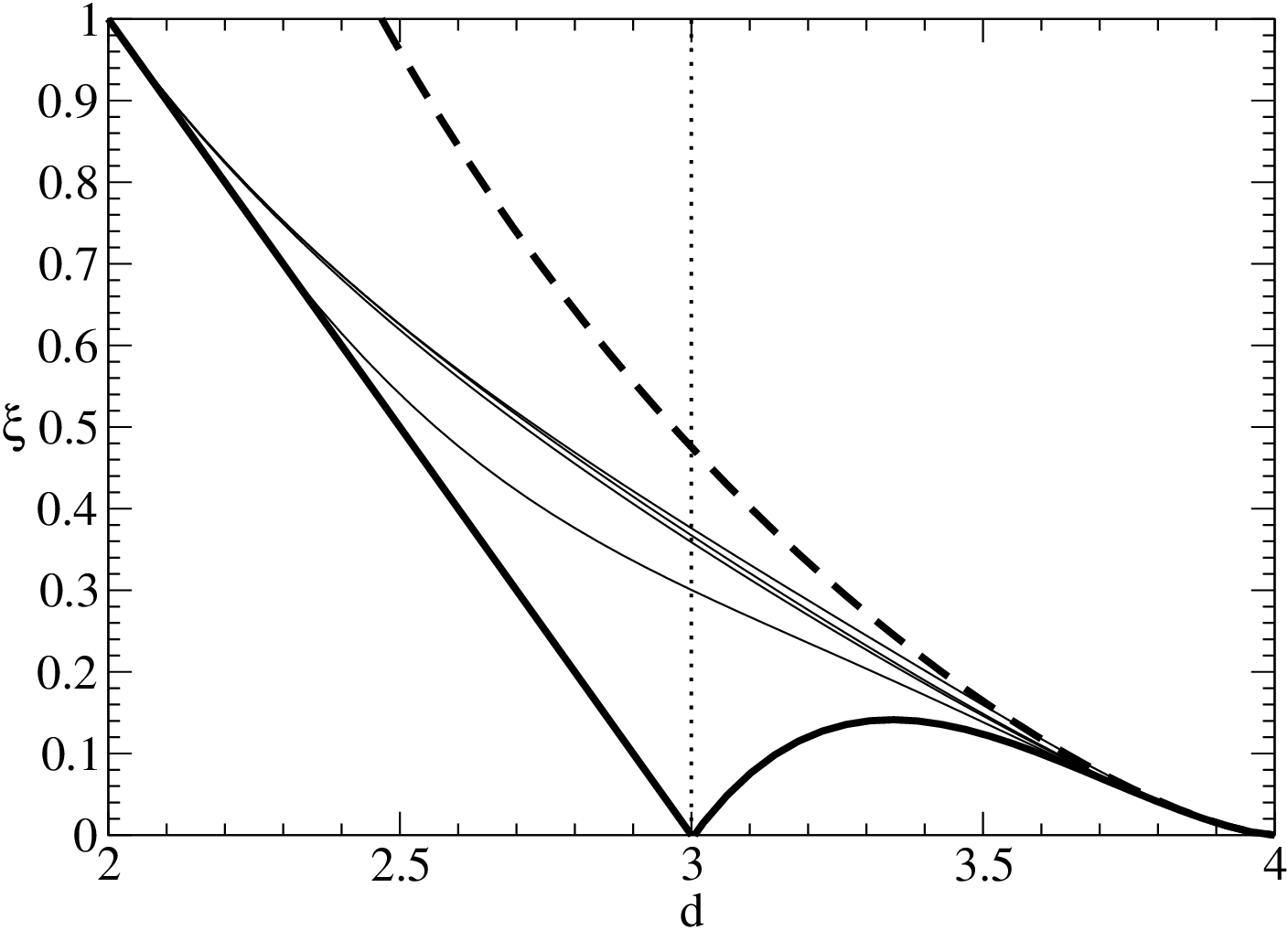}
\vspace{4pt}
\caption{%
    \label{fig:borel}
    Extrapolations of $\xi$ vs dimension $d$.  
    The thin solid lines show the result of Pad\'e-Borel extrapolations
    of type [4/0], [2/2], [3/1], and [0/4] from bottom to top.
    For comparison, the thick solid lines show simple truncations
    of the expansions about four (right) and two (left) dimensions,
    as discussed in the
    text.  The thick dashed line is the truncated $d=4-\eps$ expansion
    at NLO rather than NNLO.
    }
\end{figure}

If we naively follow the same procedure but add the information from
our NNLO coefficient, we find $\xi = 0.300$, 0.367, 0.359, and 0.376
in three dimensions
from $[4/0]$, $[3/1]$, $[2/2]$, and $[0/4]$ approximants.  We did not
find a solution for $[1/3]$.  The $[4/0]$ value is an out-lier, which
makes a certain amount of sense.  $[4/0]$ corresponds to a
simple polynomial form for $B(t)$.  This does
not allow for any singularities in
the Borel plane and so will not produce an asymptotic series
in $\eps$, whereas the large NNLO coefficient suggests that the asymptotic
nature of the expansion should not be ignored at this order.
If we focus only on the other values, they span
$\xi = 0.367\pm0.009$.  This is consistent with the NLO results,
but not an obvious improvement.  The fits of $\xi$ as a function of
$d$ are shown in Fig.\ \ref{fig:borel} and are quite similar to
the earlier fits of Ref.\ \cite{Nishida&Son2} discussed above which
did not use the NNLO
result for $d = 4-\eps$.

For comparison, the same figure also shows the result of avoiding any fancy
extrapolation but simply naively using the truncated NNLO result
$\xi = \half \eps^{1+2/d} [1 - 0.04916 \, \eps  - 0.95961 \, \eps^2]$
for $d=4-\eps$
or the corresponding NLO result
$\xi=1-\bar\eps$
of Ref.\ \cite{Nishida&Son2}
for $d=2+\bar\eps$.  It is amusing that both
of these naive extrapolations happen to give consistent values of
$\xi \simeq 0$ at $d=3$.  We imagine that this is a coincidence.

\begin{figure}[t]
\includegraphics[scale=0.50]{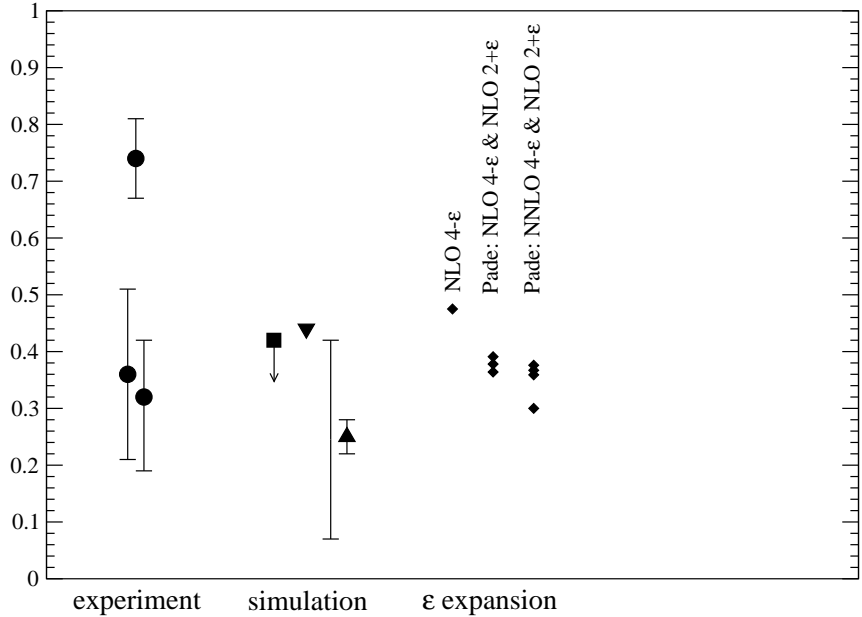}
\vspace{14pt}
\caption{%
    \label{fig:xi}
    A selection of estimates of $\xi$ in three spatial dimensions.
    Experimental values (circles) include
    0.74(7) \cite{exp_Duke}, 0.34(15) \cite{exp_ENS},
    $0.32^{+0.13}_{-0.10}$ \cite{exp_Innsbruck}, and the more recent
    0.51(4) \cite{exp_Duke2} and 0.46(5) \cite{exp_Rice}.
    Simulations include fixed-node Green's function and diffusion
    Monte Carlo upper bounds of 0.42(1) \cite{GFMC} (square).
    Other simulation methods have estimated 0.22(3) \cite{lee},
    $0.07 \le \xi \le 0.42$ \cite{schafer&lee},
    and $\approx 0.44$ \cite{bulgac} (the last two are based on 
    finite-temperature calculations). 
    The NLO $4-\eps$ value is that of Eq.\ (\ref{eq:NLO}) \cite{Nishida&Son}.
    The rest are the Pad\'e-Borel estimates discussed in
    this section, with the purely NLO result from Ref.\ \cite{Nishida&Son2}.
    }
\end{figure}

Figure \ref{fig:xi} summarizes a selection of estimates of $\xi$ from
experiment and numerical simulations, and compares them
to results obtained so far from the $\eps$ expansion.
The current Pad\'e-Borel extrapolated $\eps$ expansion results, however, need
to be taken with a significant
grain of salt because their assumptions of analytic
structure are not consistent with what we have learned in
Sec.\ \ref{sec:beyond NNLO} about higher-order corrections.
Further study will be required to develop more consistent extrapolations.


\begin{acknowledgments}

We would like to thank Yusuke Nishida for useful conversations.
One of us (P.A.) would like to thank Eric Braaten for encouragement
to become involved in this problem.
This work was supported, in part,
by the U.S. Department of Energy under Grant No.~DE-FG02-97ER41027
and Grant No.~DE-FG02-00ER41132.

\end {acknowledgments}


\appendix

\section{Summary of results by diagram type}
\label{app:summary}

\begin {align}
\makebox[7em][l]{Fig.\ \ref{fig:diags1}(a):}
\makebox[2.5em][l]{$V_1^{(0)}$} &=
    - \int_\p E_\p
  =
    \phi_0 \left( \frac{m \phi_0}{2 \pi} \right)^{d/2}
    \frac{ \Gamma(-\frac12 - \frac{d}{4}) }
           { 2_{}^{1+\frac{d}{2}} \Gamma(\frac12+\frac{d}{4}) }
  \notag\\
    &=
    \phi_0 \left( \frac{m \phi_0}{2 \pi} \right)^{d/2}
    \frac13 \Bigl\{
      1
      + \tfrac12\bigl(\tfrac73-\gamma-\ln2\bigr) \eps
  \notag\\
    & \hspace{9em}
      + \bigl[ \tfrac18 \bigl(\tfrac73-\gamma-\ln2\bigr)^2
               + \tfrac{19}{72} \bigl] \eps^2
      + O(\eps^3)
     \Bigr\}
\displaybreak[0]\\
\makebox[7em][l]{Fig.\ \ref{fig:diags1}(b):}
\makebox[2.5em][l]{$V_1^{(\mu)}$} &=
    + \mu \int_\p \frac{\eps_\p}{E_\p}
  =
    \mu \left( \frac{m \phi_0}{2 \pi} \right)^{d/2}
    \frac{ \Gamma(- \frac{d}{4}) }
           { 2^{d/2} \Gamma(\frac{d}{4}) }
  \notag\\
    &=
    - \frac{\mu}{\eps} \left( \frac{m \phi_0}{2 \pi} \right)^{d/2}
    \Bigl\{
      1
      + \half\bigl(\half-\gamma+\ln2\bigr) \eps
  \notag\\
    & \hspace{9em}
      + \bigl[ \tfrac18 \bigl(\half-\gamma+\ln2\bigr)^2
               + \tfrac1{32} \bigr] \eps^2
      + O(\eps^3)
   \Bigr\}
\displaybreak[0]\\
\makebox[7em][l]{Fig.\ \ref{fig:diags1}(c):}
\makebox[2.5em][l]{$V_1^{(\mu\mu)}$} &=
    - \half \mu^2 \phi_0^2 \int_\p \frac{1}{E_\p^3}
  =
    -
    \frac{\mu^2}{\phi_0} \left( \frac{m \phi_0}{2 \pi} \right)^{d/2}
    \frac{ \Gamma(\frac32 - \frac{d}{4}) }
         { 2_{}^{\frac{d}{2}} \Gamma(\frac12 + \frac{d}{4}) }
  \notag\\
    &=
    - \frac{\mu^2}{\phi_0} \left( \frac{m \phi_0}{2 \pi} \right)^{d/2}
   \frac12
   \bigl\{ 1 + O(\eps) \bigr\}
\displaybreak[0]\\
\makebox[7em][l]{Fig.\ \ref{fig:diags2}(a):}
\makebox[2.5em][l]{}
& \kern-2.5em\text{See Eqs.\ (\ref{eq:V20}) through (\ref{eq:C2N}).}
\notag
\displaybreak[0]\\
\makebox[7em][l]{Figs.\ \ref{fig:diags2}(b--d):}
\makebox[2.5em][l]{$V_2^{(\mu)}$} &=
    - \mu g^2 \phi_0^2 \int_{\p\q}
    \frac{(E_\p-\eps_\p)}
         {2 E_\p \, E_\q^2 \, \S_{\p\q}}
    \left( \frac{1}{E_\q} + \frac{1}{S_{\p\q}} \right)
  \notag\\
    &=
    \mu \left( \frac{m\phi_0}{2\pi} \right)^{d/2}
    [\eps \, K_2^{(\mu)} + O(\eps^2)]
\label {eq:V2mu}
\\
\makebox[7em][l]{}
\makebox[2.5em][l]{$K_2^{(\mu)}$} &\simeq -0.25835
\displaybreak[0]\\
\makebox[7em][l]{Fig.\ \ref{fig:diags3}(a):}
\makebox[2.5em][l]{}
& \kern-2.5em\text{See Eqs.\ (\ref{eq:V3cross}), (\ref{eq:V3crossK3}),
                 and (\ref{eq:K3crossN}).}
\notag
\displaybreak[0]\\
\makebox[7em][l]{Figs.\ \ref{fig:diags3}(b--d):}
\makebox[2.5em][l]{$V_3^{(\Sigma\Sigma)}$} &=
  g^4 \int_{\p\q\k}
    \frac{ (E_\p-\eps_\p) (E_\q-\eps_\q) }
         { 2 E_\p \, 2 E_\q \, (2 E_\k)^2  }
  \notag\\ & \qquad \times
    \left[
       \frac{2(E_\k-\eps_\k)^2}{\S_{\p\k}^2 \S_{\q\k}}
       - \phi_0^2 \left(
            \frac{2}{E_\k \S_{\p\k} \T_{\p\q,\k}}
            + \frac{2}{\S_{\p\k}^2 \T_{\p\q,\k}}
            + \frac{1}{E_\k \S_{\p\k} \S_{\q\k}}
         \right)
    \right]
  \notag\\
    &=
    \phi_0 \left( \frac{m\phi_0}{2\pi} \right)^{d/2}
    \bigl[ \eps^2 \, K_3^{(\Sigma\Sigma)} + O(\eps^3) \bigr]
\label {eq:V3sigsig}
\\
\makebox[7em][l]{}
\makebox[2.5em][l]{$K_3^{(\Sigma\Sigma)}$} &\simeq -0.03046
\displaybreak[0]\\
\makebox[7em][l]{Fig.\ \ref{fig:diags3}(e):}
\makebox[2.5em][l]{$V_3^{(\widetilde\Pi\widetilde\Pi)}$} &=
    - g^4 \phi_0^4 \int_{\p\q\k}
    \frac{ 1 }
         { 2 E_\p \, 2 E_\q \, 2 E_\k \, 2 E_\l \, \S_{\p\k} \S_{\q\l}}
       \left[
         \frac{1}{\half\eps_{\p+\k}}
          + \frac{1}{\E_{\p\q\k\l}}
       \right]
  \notag\\
    &=
    \phi_0 \left( \frac{m\phi_0}{2\pi} \right)^{d/2}
    \bigl[ \eps^2 K_3^{(\widetilde\Pi\widetilde\Pi)} + O(\eps^3) \bigr]
\label {eq:V3pipiTilde}
\\
\makebox[7em][l]{}
\makebox[2.5em][l]{$K_3^{(\widetilde\Pi\widetilde\Pi)}$} &\simeq -0.31080
\displaybreak[0]\\
\makebox[7em][l]{Figs.\ \ref{fig:diags3}(f--h):}
\makebox[2.5em][l]{}
& \kern-2.5em\text{See Eqs.\ (\ref{eq:V3pipi}), (\ref{eq:K3pipiN})
                 and (\ref{eq:C2logS}).}
\notag
\end {align}
We have used the shorthand notation of Eq.\ (\ref{eq:shorthand}).
In the momentum integrals shown in Eqs.\ (\ref{eq:V2mu}) 
and (\ref{eq:V3sigsig}),
we have used symmetry to simplify the
expressions.  For expressions that correspond to simply doing the
frequency integrals of these diagrams without using such symmetry,
simply symmetrize the integrands under $\p \leftrightarrow \q$.


\section{What happened to \boldmath$\theta(\mu-\eps_\p)$?}
\label{app:theta}

In this appendix, we give a quick, illustrative example of
the effect of $\phi_0$ on what otherwise would be a Fermi-sea
step-function $\theta(\mu-\eps_\p)$.  The example will be the
mean-field theory result for the fermion number density $n$
in the presence of both a non-negligible chemical potential $\mu$
and a condensate $\phi_0$.
In terms of diagrams, this corresponds to the one-loop fermion
diagram of Fig.\ \ref{fig:diags1}(b) if we take the cross to represent the
fermionic number operator and (unlike the rest of this paper)
include the chemical potential $\mu$ in the fermion propagator
rather than treating it as a perturbation.  It will be clearer
if we give a regularization-independent version of the result
rather than continuing to work in dimensional regularization.
In terms of diagram evaluation, this can be achieved by subtracting
the vacuum contribution ($\mu=\phi=0$) to $n$ before
doing any integrals.  The result, after frequency integration, is
then the standard mean-field formula for the number equation%
\footnote{
  For comparison, the result to leading order in $\mu$ in
  dimensional regularization could be obtained by applying
  $n = -\partial V/\partial\mu$ to the result
  $V_1 = - \int_\p E_\p^\mu$ discussed in section \ref{sec:remaining},
  giving $n = - \int_\p \eps_\p^\mu/E_\p^\mu$
  in dimensional regularization.
  This appears to differ from
  Eq.\ (\ref{eq:nmeanfield}) by $\int_\p 1$, but $\int_\p 1$ vanishes
  in dimensional regularization.
}
\begin {equation}
  n = \int_\p \left( 1 - \frac{\eps_\p^\mu}{E_\p^\mu} \right)
    = \int_\p \left( 1 -
           \frac{(\eps_\p-\mu)}{[(\eps_\p-\mu)^2+\phi_0^2]^{1/2}}
           \right) .
\label {eq:nmeanfield}
\end {equation}
In the limit $\phi_0 \to 0$, the above formula gives the usual result
for a free fermion gas:
\begin {equation}
  n = 2 \int_\p \theta(\mu - \eps_\p) .
\end {equation}
It would be problematical to treat $\mu$ perturbatively in such an
expression.  However, in the opposite limit
$\mu \ll \phi_0$ relevant to this paper, there is no
obstruction to treating $\mu$ perturbatively.
To leading order in $\mu$,
\begin {equation}
  n = \int_\p \left( 1 - \frac{\eps_\p}{E_\p} \right) .
\end {equation}
We thus find fractions involving energies rather than
$\theta$ functions, similar to results we have derived for other
diagrams in this paper.




\end{document}